\documentclass[twocolumn,showpacs,preprintnumbers,amsmath,amssymb,natbib,noshowpacs]{revtex4-1}
\usepackage{graphicx}% Include figure files
\usepackage{dcolumn}% Align table columns on decimal point
\usepackage{bm}% bold math
\usepackage{color}
\usepackage{hyperref}
\usepackage{amssymb}
\usepackage{amsmath,empheq}
\DeclareMathOperator{\sech}{sech}
\usepackage{tikz}
\makeatletter
\def\l@subsection#1#2{}
\def\l@subsubsection#1#2{}
\makeatother
\begin{document}

\preprint{OU-HET-908}

\title{Boundary Conditions of Weyl Semimetals}% Force line breaks with \\

\author{Koji Hashimoto}
\email{koji@phys.sci.osaka-u.ac.jp}
 \affiliation{Department of Physics, Osaka University,
 Toyonaka, Osaka 560-0043, Japan.}%Lines break automatically or can be forced with \\
\author{Taro Kimura}%
 \email{taro.kimura@keio.jp}
\affiliation{%
Department of Physics, Keio University, Kanagawa
 223-8521, Japan.}
\author{Xi Wu}
\email{wuxi@het.phys.sci.osaka-u.ac.jp}
\affiliation{Department of Physics, Osaka University,
 Toyonaka, Osaka 560-0043, Japan.}
%\affiliation{


\begin{abstract}
We find that generic boundary conditions of the Weyl semimetal is dictated by only a single real parameter,
in the continuum limit.
We determine how the energy dispersions (the Fermi arcs) and the wave functions of edge states 
depend on this parameter. Lattice models are found to be consistent with our generic observation.
Furthermore, the enhanced parameter space of the boundary condition is shown to support a 
novel topological number.
\end{abstract}

\maketitle

%\tableofcontents

%%%%%%%%%%%%%%%%%%%%%%%%%%%%%%%%%%%%%

\section{Introduction}\label{sec:intro}

One of the important aspects of topological phases \cite{hasan2010colloquium,qi2011topological}
is that they bridge condensed matter physics and particle physics. Familiar and important concepts in particle physics, such as topological charges, quantum anomalies
and relativity, are completely translated to condensed matter physics, and their experimental realization
flourishes and verifies the concept quite nontrivially. The interplay between the condensed matter physics and the particle physics is further expected to provide developments of these fields.

From the theoretical side, topological phases are classified by dimensions and discrete symmetries~\cite{Schnyder:2008tya,Kitaev:2009mg}. The symmetries
are important for classification, as they are robust against deformations. In particular, at microscopic levels
materials have lots of ways for their Hamiltonian to be deformed. However, if one is based on the topological
properties which are determined by the discrete symmetries, the classification and the resultant phenomena are robust. This works even at a continuum limit where detailed structure of lattice Hamiltonian disappears.
The outcome of the topological phases is the existence of gapless edge mode, as a consequence of
the renowned bulk-edge correspondence \cite{Jackiw:1975fn,hatsugai1993chern,Wen:2004ym}.

Particle physics Lagrangians are written based on symmetries, too. Normally, particle physics 
Lagrangians have all possible terms allowed by the required symmetries, in the continuum limit.
Then a natural question arises: what are all possible boundary conditions?
This question is particularly important in view of the bulk-edge correspondence, because
one typically uses open boundary conditions in the study of topological phases.
%In view of the bulk-edge correspondence, a natural question arises: one typically uses %periodic and 
%open boundary conditions in the study of topological phases, but what is all possible boundary conditions?

Although to answer this question at the level of listing all possible lattice Hamiltonians with boundaries is 
far beyond our knowledge, we could approach the answer in the continuum limit. The continuum Hamiltonian
is much easier to deal with, and we can ask how general the Hamiltonian at our hand is. Furthermore,
it is expected to capture the generic feature of all lattice systems which share the same discrete symmetries
as the continuum Hamiltonian. 

In this paper, we focus on the 3D Weyl semimetal, which has been recently observed in experiments~\cite{Xu:2015Science,Huang:2015NC,Weng2015:PRX} after theoretical predictions~\cite{Murakami:2007bx,Murakami:2007NJP,Wan:2011PRB,Yang:2011PRB,Burkov2011:PRL,Xu:2011dn,Burkov2011:PRB}, and study the most generic boundary condition in the continuum limit.
We will
see how a generic surface term in the Lagrangian can affect the edge states of a Weyl fermion, in 
a Hamiltonian language -- in other words, we will see how a generic modification of the 
boundary conditions give change in edge dispersions. 
Surely the existence of edge states for a given Weyl semimetal is explained by 
the topological number of bulk theory, but how the boundary conditions are related with the edge states, 
or so-called Fermi arcs, have not been studied well in the literature.\footnote{For generic boundary conditions 
for topological insulators, see \cite{isaev2011bulk,Enaldiev:2015JETP}. For transitions from Weyl semimetals to topological insulators, see \cite{okugawa2014dispersion}.} 
To look at genericity under a given symmetry is a particle theory standpoint, which could serve as another new interplay between
the two fields.

Our study is divided into two parts, the study of the continuum theory and a verification by lattice models. 
On the continuum theory side, we start with the standard Hamiltonian with a single Weyl cone, 
and consider the most general boundary condition, with resultant edge states.
In particular, we are careful with allowed number of parameters.
We also study the dimensional reduction to 2D, at which topological insulators of class A 
can be studied. 
On the lattice model side, a graphene in 2D and a square lattice in 3D are considered. 

Our main findings are summarized as follows:
\begin{itemize}
\item The generic boundary condition of the 3D Weyl semimetal is dictated only by 
a single real parameter. 
At low energy, all edge states are labeled by the parameter.
\item 
The parameter describes how the edge dispersion is attached to the Weyl node.
\item Results of the lattice models are perfectly consistent with the continuum theory in the vicinity of the Weyl point.
\end{itemize}

In the course of our study, we also find novel properties of the Weyl semimetals:
\begin{itemize}
\item Our results of the edge states are all consistent with the bulk-edge correspondence.
In particular we propose how to count the edge states for the Weyl semimetals.
\item A new topological structure emerges in the enhanced parameter space spanned
by the boundary parameter and the two conserved momentum. 
The edge state carries the topological charge.
\end{itemize}

Our paper is organized as follows. 
In Sec.~\ref{sec:bc_3DWeyl}, we study the 3D Weyl semimetals and their generic boundary conditions in the continuum limit.
We derive the relations between energy dispersions, wave functions of edge states and the boundary conditions. We study also a dimensional reduction to 2D topological insulators. 
In Sec.~\ref{sec:lattice}, we investigate lattice models and show the consistency with the results 
obtained in the continuum limit.
In Sec.~\ref{sec:bec} we check the bulk-edge correspondence. We further
propose how to count the edge states of the Weyl semimetals.
In Sec.~\ref{sec:new_top_num} we find a new topological structure of the edge states
in the enhanced parameter space of the boundary conditions.
In the last section, we summarize our paper.
In Appendix, we provide two examples of generalization: treatment of generic momentum-dependent
boundary conditions, and the case with two parallel boundaries, in the continuum limit of the model of the Weyl semimetals.

%%%%%%%%%%%%%%%%%%%%%%%%%%%%%%%%%%%%%

\section{Generic boundaries of 3D Weyl semimetals at continuum}\label{sec:bc_3DWeyl}

%%%%%%%%%%%%%%%%%%%%%%%%%%%%%%%%%%%%%

We start with a generic Weyl semimetal in 3 spatial dimensions.
Near the Weyl point, the Hamiltonian in the continuum limit is generically given by 
\begin{align}\label{Ham}
	\mathcal{H}=p_1\sigma_1+p_2\sigma_2+p_3\sigma_3.
\end{align}
Here the Weyl point sits at the origin of the 3-dimensional momentum space.
We are interested in the most generic boundary conditions and the spectra, of this system.
The states are energy eigenstates, subject to the following boundary condition:
%
%Our starting point is a 3+1 dimensional Hamiltonian describing a two-state system with a generic 
%consideration of matrix $M$ in the boundary condition:
\begin{empheq}[left=\empheqlbrace]{align}
		&\label{HE} \mathcal{H}\psi=\epsilon\psi \\ 
		\label{BC} &(M+1)\psi\Big|_{x^3=0}=0,
\end{empheq}
where we put a boundary at $x^3=0$, 
so the material is in the spatial region $x^3\geq 0$.
A real constant
$\epsilon$ is the energy eigenvalue, and 
$M$ is a generic $2\times 2$ complex constant matrix. (See App.~\ref{sec:momentum} for 
generic momentum-dependent boundary conditions.)

%with 
%\begin{align}\label{Ham}
%	\mathcal{H}=p_1\sigma_1+p_2\sigma_2+p_3\sigma_3.
%\end{align}
The Hamiltonian can effectively describe a two-band system that has a degeneracy 
at the Weyl point with a definite chirality. 
It captures the topological nature around the Weyl point, and is equivalent to the Hamiltonian of a Weyl fermion in 3 spatial dimensions. 
%Two components of  spinor $\psi$ are two states in the two different bands. 
The boundary condition needs to be of the form (\ref{BC}) since at the boundary two components of $\psi$ are related to each other through the boundary condition.
%here connects these two states on the boundary. Here we omit spin degrees of freedom. 
The matrix $M$ reflects the arbitrariness in the choice of the boundary condition; One can think of
infinitely many kinds of boundaries, starting from just a slicing of the material, to putting various chemical layers on top of the boundary surfaces, such as hydrogen/nitrogen termination or oxidization.

We solve the equations above, which give energy dispersion relations and wave functions.
We find that the most generic boundary condition \eqref{BC} %specified by the arbitrary matrix $M$
is parameterized only by a single real constant $\theta_+ \in [0,\pi)$.
% and Berry connections for edge modes. We will see that a nontrivial topological number appears.

%%%%%%%%%%%%%%%%%%%%%%%%
\subsection{Parameterizing generic boundary condition}
\label{sec:bc_cont}
%%%%%%%%%%%%%%%%%%%%%%%%
In this subsection we explore all the requirements for the boundary condition matrix $M$
in (\ref{BC})
and parametrize it. It turns out that $M$ 
is parameterized only by two real parameters, and 
furthermore, the boundary condition (\ref{BC}) needs only one of them. 
These result from a self-conjugacy of the Hamiltonian \eqref{Ham} and the requirement 
that $M$ needs to have the eigenvalue $-1$: the combination
$M+1$ needs to have a zero eigenvalue, for a non-trivial solution satisfying the boundary condition \eqref{BC}
to exist.

%%%%%%%%%%%%%%%%%%%%%%%%%%%%%
\subsubsection{Self-conjugacy of the Hamiltonian}

In general, $M$ is a $2\times 2$ complex matrix and has 4 complex 
degrees of freedom (d.o.f.), that is, 8 real d.o.f. are there. We parametrize it as
\begin{align}
	M=a_0 \mathbf{1}_2 + a_i \sigma_i
\end{align}
with complex coefficients $a_0=A_0+iB_0$ and $a_i = A_i + iB_i \in \mathbb{C}$, where
$i=1,2,3$.%and $\sigma_0\equiv \mathbf{1}_2$.

We need our Hamiltonian \eqref{Ham} to be self-conjugate, which 
gives a constraint on the boundary condition.
See also \cite{Witten:2015aoa}.
The self-conjugacy condition of Hamiltonian is
\begin{align}
	\langle \mathcal{H}\psi_1 | \psi_2 \rangle= \langle \psi_1 | \mathcal{H} \psi_2 \rangle,
\end{align}
for arbitrary normalizable $\psi_1$ and $\psi_2$.
When we explicitly write the two inner product above as an integration, 
we find a surface difference between the right hand side and the left hand side, which must vanish:
\begin{align} 
	[(\sigma_3\psi_1)^{\dagger}\psi_2]|_{x_3=0}=0.
\end{align} 
Applying the boundary condition (\ref{BC}) onto this equation gives 
\begin{align*}
	&[(\sigma_3\psi_1)^{\dagger}\psi_2]|_{x_3=0}
	\\
	=&-\frac{1}{2}[(\sigma_3M\psi_1)^{\dagger}\psi_2]|_{x_3=0}
	-\frac{1}{2}[(\sigma_3\psi_1)^{\dagger}M\psi_2]|_{x_3=0} 
	\\
	=&-\frac{1}{2}[((\sigma_3M+M^{\dagger}\sigma_3)\psi_1)^{\dagger}\psi_2]|_{x_3=0}=0
\end{align*}
which is satisfied for any choice of $\psi_1$ and $\psi_2$ only when 
\begin{align}\label{MSSM}
	M^{\dagger}\sigma_3=-\sigma_3M.
\end{align}
This is required by the self-conjugacy of the Hamiltonian.
Equation \eqref{MSSM} removes four real d.o.f of $M$: 
\begin{align}
	0&=M^{\dagger}\sigma_3+\sigma_3M
	\nonumber \\
	&=a_I^{*}\sigma^I\sigma^3+\sigma^3\sigma^Ia_I \nonumber
	\\
	&=2A_0\sigma_3+2A_2\mathbf{1}_2+2iB_1\sigma_2-2iB_2\sigma_1,
\end{align}
and results in 
 \begin{align} \label{AABB}
 	A_0=A_3=B_1=B_2=0. 
 \end{align}
So we are left with the boundary condition matrix with four real parameters,
\begin{align}
M = A_1 \sigma_1 + A_2 \sigma_2 + i B_0 \mathbf{1}_2 + i B_3 \sigma_3 \, .
\end{align}

%%%%%%
\subsubsection{Eigenvalues of $M$}

The boundary condition (\ref{BC}) can be regarded as an eigen equation for matrix $M$. Substituting equation \eqref{AABB} into the determinant of (\ref{BC})
\begin{align}
	\text{det}(M-\lambda)=0,
\end{align}
 we get
\begin{align}
	\lambda_{\pm}=iB_0\pm \sqrt{A_1^2+A_2^2-B_3^2}.
\end{align}
The boundary condition requires $M$ to have a real eigenvalue $-1$, which gives two constraints:
\begin{align}
	B_0=0,
	\\
	\pm \sqrt{A_1^2+A_2^2-B_3^2}=-1.
\end{align}
As a result, the generic boundary condition matrix should be written as
\begin{align}
	M=A_1\sigma_1+A_2\sigma_2+iB_3\sigma_3,
\quad	A_1^2+A_2^2-B_3^2=1.
\end{align}
Consequently, $M$ is parametrized by two real parameters.
We can choose
\begin{empheq}[left=\empheqlbrace]{align}
		\nonumber &A_1=\cos\theta\cosh\chi\\  
		&A_2=\sin\theta\cosh\chi \\
		\nonumber &B_3=\sinh\chi
\end{empheq}
for parameterizing the matrix, with which 
the boundary condition can be written as 
\begin{align}\label{bc1}
	\left(\begin{array}{cc}
	\sech{\chi}+i\tanh{\chi} & \cos{\theta}-i\sin{\theta} 
	\\
	\cos{\theta}+i\sin{\theta} & \sech{\chi}-i\tanh{\chi}
	\end{array}\right)
	\psi\Big|_{x^3=0}=0.
\end{align}
Defining $\cos{\theta'}=\sech{\chi}$ and $\sin{\theta'}=\tanh{\chi}$ and changing variables: 
\begin{align*}
	\theta'&=\theta_++\theta_-
	\\
	\theta&=\theta_+-\theta_-
\end{align*}
we find that the equation  \eqref{bc1} becomes
\begin{align}
	\left(\begin{array}{cc}
	e^{i\theta'} & e^{-i\theta}  
	\\ 
	e^{i\theta} & e^{-i\theta'} 
	\end{array}\right)
	\psi\Big|_{x^3=0}=0.
\end{align}
Noting a relation 
\begin{align}
\left(\begin{array}{cc}
	e^{i\theta'} & e^{-i\theta}  
	\\ 
	e^{i\theta} & e^{-i\theta'} 
	\end{array}\right)
	=
	\left(\begin{array}{c}
	e^{i\theta'} \\
	e^{i\theta} 
	\end{array}\right)
	\left(\begin{array}{cc}1 & e^{-2i\theta_+} 
	\end{array}\right),
\end{align}
the boundary condition is recast to the following simple form
\begin{align}\label{bc'}
	\left(\begin{array}{cc}1 & e^{-2i\theta_+} 
	\end{array}\right)
	\psi\Big|_{x^3=0}=0.
\end{align}
It is surprising that
actually for the boundary condition we need only a single real parameter $\theta_+$. 
So we conclude that the most generic boundary condition is just dictated by a single real
parameter.

This equation (\ref{bc'}) tells us further that, at the boundary, two components of the fermion 
need to have the identical magnitude, and the relative phase between them is determined 
by $\theta_+$. This is true for the edge modes as well as the bulk modes. 

For our later purpose, we determine here the range of the parameter $\theta_+$. 
First, from the definition of $\theta$ and $\theta'$, we find that 
they live on a region
\begin{align}
0<\theta\leq 2\pi, \quad 0<\theta' \leq \pi \, .
\end{align}
Note that $\sin\theta'=\sech\chi\geq0$. Then one notices that the region
can be equally covered by 
\begin{align}
0<\theta\leq 2\pi, \quad	0 \leq \theta+\theta' \leq 2\pi.
\end{align}
Resultantly, the smallest necessary region for $\theta_+$ is given by
\begin{align}
0\leq \theta_+\leq \pi.
\end{align}
$\theta_+=0$ and $\theta_+=\pi$ are the same configuration up to some adjustment of
$\theta_-$ which does not appear in the boundary condition itself.

%%%%%%%%%%%%%

%%%%%%%%%%%%%%%%%%%%%%%%%%%%%%%%%%%%%

\subsection{Lagrangian formulation}

Since Lagrangian formulation sometimes works easier, here we present an equivalent
Lagrangian formulation of what we have seen in terms of the Hamiltonian.
In fact we find that the condition \eqref{MSSM} shows up naturally in the Lagrangian formulation.
Let us describe a generic and consistent boundary condition of a Weyl semimetal
in 1+3 spacetime dimensions.
Metric convention is chosen as $\eta_{\mu\nu}=\mbox{diag}(+,-,-,-)_{\mu\nu}$. 
%and the gamma
%matrices are with $\{\gamma^\mu,\gamma^\nu\}=2\eta^{\mu\nu}$  ($\mu,\nu=0,1,3$). 
%Note
%that our spacetime is spanned by $(x^0,x^1,x^3)$, following the convention of
%our dimensional reduction.

The bulk Lagrangian (for a right-handed Weyl fermion) 
is written as % (see \cite{vanNieuwenhuizen:1996tv} for a 1+3-dimensional case)
\begin{align}
{\cal L} = \frac{i}{2} \psi^\dagger \sigma^\mu (\overrightarrow{\partial}_\mu 
 - \overleftarrow{\partial}_\mu)
%(\gamma^\mu \partial_\mu + m) 
\psi
\label{eq:Lag2}
\end{align}
where $\sigma^\mu=({\bf 1}_2, \sigma_1,\sigma_2,\sigma_3)$.
%with $\bar{\psi}\equiv \psi^\dagger i \gamma^0$.
The Dirac equation is
\begin{align}
\sigma^\mu \partial_\mu  \psi
=0
\end{align}
which can be rewritten as
\begin{align}
\left[i\partial_0 +i\sigma_i \partial_i\right]\psi
=0
\end{align}
where $i=1,2,3$.
So the Hamiltonian is $i\partial_0 = {\cal H}$, 
\begin{align}
{\cal H}  = p_1 \sigma_1 + p_2 \sigma_2 + p_3 \sigma_3,
\label{eq:hamil2}
\end{align}
which is the standard Hamiltonian of the Weyl semimetal near the Weyl cone.

Let us introduce a surface term in the Lagrangian, for deriving the boundary condition. 
The total action is
\begin{align}
S = \int_{x^3\geq 0} \!\!\!\!d^3x \; 
\frac{i}{2} \psi^\dagger \sigma^\mu (\overrightarrow{\partial}_\mu 
 - \overleftarrow{\partial}_\mu)
\psi
 + \frac12 \int_{x^3=0} \! d^2x \;
\psi^\dagger N \psi \, .
\end{align}
The first term is the Weyl Lagrangian. 
The second integral is with a Hermitian matrix $N$.
A variation $\psi\to\psi + \delta \psi$ and $\psi^\dagger \to \psi^\dagger+\delta\psi^\dagger$ 
provides equations at the surface $x^3=0$ as
\begin{align}
\left[ - i\psi^\dagger \sigma_3 + \psi^\dagger N \right]\delta\psi= 0 , \quad 
 \delta\psi^\dagger
 \left[ i\sigma_3 \psi + N\psi \right] = 0 .
\end{align}
For this to be valid for arbitrary $\delta\psi$ and $\delta\bar\psi$, we find
\begin{align}
- i\psi^\dagger \sigma_3 + \psi^\dagger N = 0 , \quad 
i\sigma_3 \psi + N\psi = 0 
\label{boudagger}
\end{align}
at the boundary $x^3=0$.
These two equations are complex-conjugate to each other. If we write
$N =  i \sigma_3 M$, then
\begin{align}
(M+1) \psi\Big|_{x^3=0} = 0.
\end{align}
The Hermiticity condition $N=N^\dagger$ is now written by $M$ as
\begin{align}
M^\dagger \sigma_3 + \sigma_3 M = 0
\label{M22M}
\end{align}
which is found to be equivalent to the Hamiltonian conjugacy constraint on $M$, (\ref{MSSM}).
Note that this condition follows from the Hermiticity of $N$, that is, the
Hermiticity of the surface Lagrangian.

So, in the end, we found that the boundary condition is dictated by
a boundary ``mass'' term with a Hermitian matrix $N$,
\begin{align}
\frac12 \int d^3x \; \delta(x^3) \; \psi^\dagger N \psi \, .
\end{align}
The most generic boundary condition is given by a constant 
Hermitian matrix $N$, and that is physically natural.

%%%%%%%%%%%%%%%%%%%%%%%%%%%%%%%%%%%%%
%%%%%%%%%%%%%%%%%%%%%%%%
\subsection{Generic edge modes and dispersions} \label{sec:generic}
%%%%%%%%%%%%%%%%%%%%%%%%
Since the Weyl fermion possesses a topological number, we expect the existence
of the edge-localized modes when we introduce a boundary $x^3=0$.
In this subsection we look for edge mode solution of the energy eigenvalue problem. With the most generic boundary condition \eqref{bc'} we obtained, we find a dispersion relation and the 
wave function of the edge mode, which are completely specified by $p_1, p_2$ and $\theta_+$.

\subsubsection{Solving eigenstate equation}

Now we look for edge mode solution to eigenvalue equation \eqref{HE}. With an explicit two-component notation
\begin{align}
	\psi=
	\left(\begin{array}{c} 
		\xi \\
		\eta
	\end{array}\right),
\end{align}
the eigenstate equation \eqref{HE} can be written as
\begin{align}\label{HE'}
	\left(\begin{array}{cc}
	-i\partial_3-\epsilon & p_1-ip_2 
	\\
	p_1+ip_2  & i\partial_3-\epsilon
	\end{array}\right)
	\left(\begin{array}{c}\xi \\\eta\end{array}\right)
	=0.
\end{align}
This equation can be reorganized into two independent second-order differential equations:
\begin{align}
%	\left\{\begin{array}{c}
%	(p_1^2+p_2^2-\epsilon^2-\partial_3^2)\xi=0 
%	\\
%	(p_1^2+p_2^2-\epsilon^2-\partial_3^2)\eta=0.
%	\end{array}\right.
 \left(p_1^2+p_2^2-\epsilon^2-\partial_3^2\right)
 \begin{pmatrix}
  \xi \\ \eta
 \end{pmatrix} = 0 \, .
\end{align}
We look for the modes localized at the boundary. For the edge modes, we need 
\begin{align}\label{al}
	\alpha^2\equiv p_1^2+p_2^2-\epsilon^2>0,
\end{align}
then the corresponding solutions required by the normalizability are
\begin{align} \label{EM}
%	\left\{\begin{array}{c}
%	\xi=\xi_0\text{exp}(-\alpha(\epsilon) x^3) \\
%	\eta= \eta _0\text{exp}(-\alpha(\epsilon) x^3),
%	\end{array}\right.
 \begin{pmatrix}
  \xi \\ \eta
 \end{pmatrix}
 = e^{- \alpha(\epsilon) x^3}
 \begin{pmatrix}
  \xi_0 \\ \eta_0
 \end{pmatrix}
 \, ,
\end{align} 
where $\xi_0$ and $\eta_0$ have no dependence on $x^3$.
These are the edge modes, and in the following we determine the dispersion $\epsilon(p_1,p_2)$
and the relation between the components $\xi_0$ and $\eta_0$.

%%%%
\subsubsection{Dispersion relation}
We combine the results from eigenvalue equation \eqref{HE} and boundary condition \eqref{BC} for edge eigen modes.
Substituting equations \eqref{al} and \eqref{EM} into equation \eqref{HE'}, we get one independent equation:
\begin{align}\label{HE"}
	(i\alpha-\epsilon)\xi_0+ (p_1-ip_2)\eta_0=0.
\end{align}
Writing the boundary condition \eqref{bc'} and equation \eqref{HE"} together, we have 
\begin{align}\label{xi0eta0}
	\left(\begin{array}{cc}
	i\alpha-\epsilon & p_1-ip_2 
	\\
	1 & e^{-2i\theta_+}
	\end{array}\right)
	\left(\begin{array}{c}\xi_0 \\\eta_0\end{array}\right)
	=0.
\end{align}
This matrix equation 
contains the essence of the eigenvalue equation \eqref{HE} and 
the boundary condition \eqref{BC} for edge eigen modes. The vanishing determinant condition of 
\eqref{xi0eta0} gives 
\begin{align}\label{*}
	e^{-2i\theta_+}(i\alpha-\epsilon)=p_1-ip_2.
\end{align}
We move the $\epsilon$ term to the right, square both sides and cancel $p_1-ip_2$, finding 
the relation:
\begin{align}\label{EDR}
	\epsilon=-p_1\cos{2\theta_+}-p_2\sin{2\theta_+}.
\end{align}
This is the dispersion relation of the edge states.
It is linear with respect to $p_1$ and $p_2$, and {\em speed of light} is now anisotropic. 

Substituting \eqref{EDR} back into equation \eqref{*} we also find
\begin{align}
	\alpha=p_1\sin{2\theta_+}-p_2\cos{2\theta_+}.
\end{align}
We write above two equations in a compact way:
\begin{align}
	\left(\begin{array}{c}\epsilon \\\alpha\end{array}\right)
	=
	-\left(\begin{array}{cc}\cos{2\theta_+} & \sin{2\theta_+} \\ -\sin{2\theta_+} & \cos{2\theta_+}\end{array}\right)
	\left(\begin{array}{c}p_1  \\p_2\end{array}\right).
	\label{rotea}
\end{align}
%It is easy to see that energy eigenvalue depends only on $2\theta_+$.
Interestingly, (\ref{rotea}) shows that
what the boundary does is only rotating the momenta $(p_1,p_2)$ into $(\epsilon,\alpha)$, the energy 
 and the inverse of edge mode decay width (penetration depth). 
For fixed $p_1$ and $p_2$, we can regard the pair $(\epsilon, \alpha)$ as a vector rotating around the origin
by $2\theta_+$. 
%two components of a vector after a rotation of angle $2\theta_++\pi$. So w
When the absolute value of $\epsilon$ becomes large, $\alpha$ becomes small, then the 
penetration depth is large.
On the other hand, when the absolute value of $\epsilon$ becomes small, $\alpha$ becomes large and then the 
penetration depth is small.
This coincides with the intuition that the wave function penetration measured from the location of the boundary
increases for larger energy of the edge mode.

\begin{figure}[h!]
	\begin{center}
	 \includegraphics[scale=0.25]{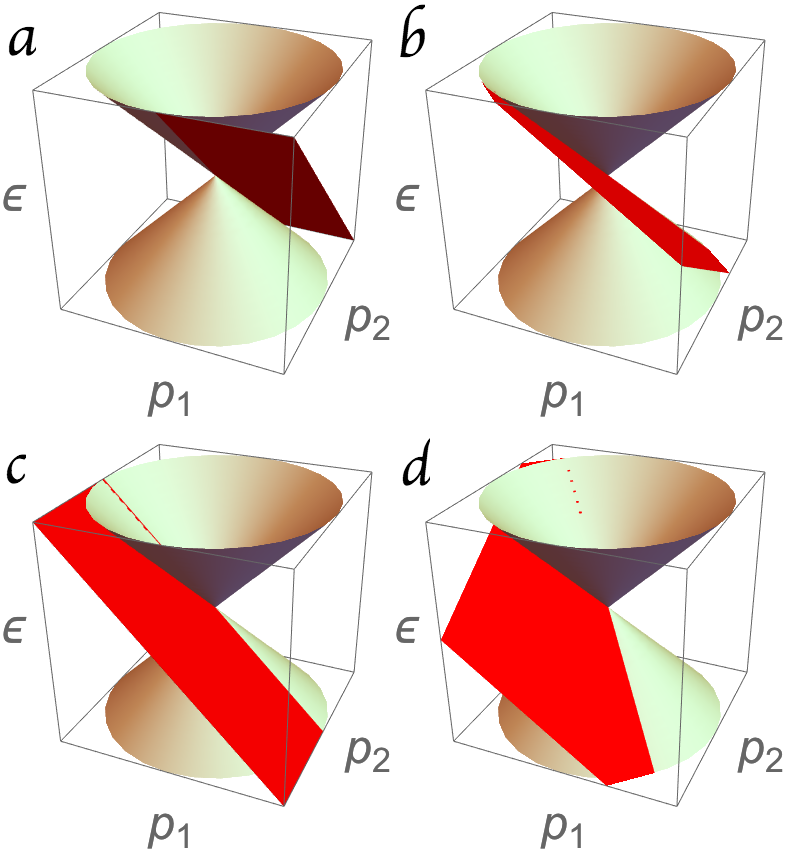}
	 \includegraphics[scale=0.25]{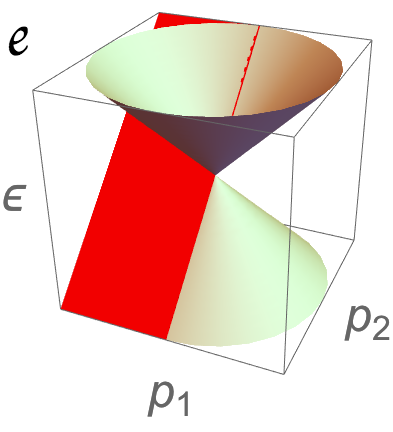}	 
	 \caption{Figures $a$, $b$ $c$, $d$, $e$ respectively represent the energy dispersions of the bulk states and the edges states, for $2\theta_+=\pi/2, \pi/4, 0, -\pi/4, -\pi/2$.}
	 \label{Dispersion1}
          \end{center}
\end{figure}

Plotting the dispersion relation, we actually see in Fig.~\ref{Dispersion1} that the edge dispersion is rotated against the $(p_1,p_2)$ axes by the change of the boundary parameter $\theta_+$.

%%%%
\subsubsection{Wave function of edge modes}
Let us finally write the wave function of the edge states.
We have already used up most of the information and are left with normalization condition only, with 
which we can determine the wave function completely. Substituting \eqref{EM} to 
the normalization condition
\begin{align}
	\int^\infty_0 dx^3~\psi^{\dagger} \psi=1,
\end{align}
we obtain a constraint
\begin{align}\label{NC}
	|\xi_0|^2+|\eta_0|^2=2\alpha.
\end{align}
With the second equation of \eqref{xi0eta0}, the boundary condition, we can see that the two components should have the same magnitude with a difference of their phases. 
Combined with \eqref{NC}, they are determined up to an irrelevant overall phase:
\begin{align}
%	\left\{\begin{array}{c}
%	\xi_0=\sqrt{\alpha} e^{-2i\theta_+}
%	\\
%	\eta_0=-\sqrt{\alpha}
%	\end{array}\right..
 \begin{pmatrix}
  \xi_0 \\ \eta_0
 \end{pmatrix}
 = \sqrt{\alpha}
 \begin{pmatrix}
  e^{-2i\theta_+} \\ -1
 \end{pmatrix}
 \, .
\end{align}
So the general edge mode wave function is 
\begin{align}
\label{edgegeneral}
	\psi(x^3)&=\sqrt{\alpha}~\text{exp}(-\alpha x^3)
	\left(\begin{array}{c} e^{-2i\theta_+} \\ -1
	\end{array}\right),
	\\ \alpha&=p_1\sin{2\theta_+}-p_2\cos{2\theta_+}. \nonumber
\end{align}

\begin{figure*}[t] 
	\begin{center}
	 \includegraphics[scale=0.25]{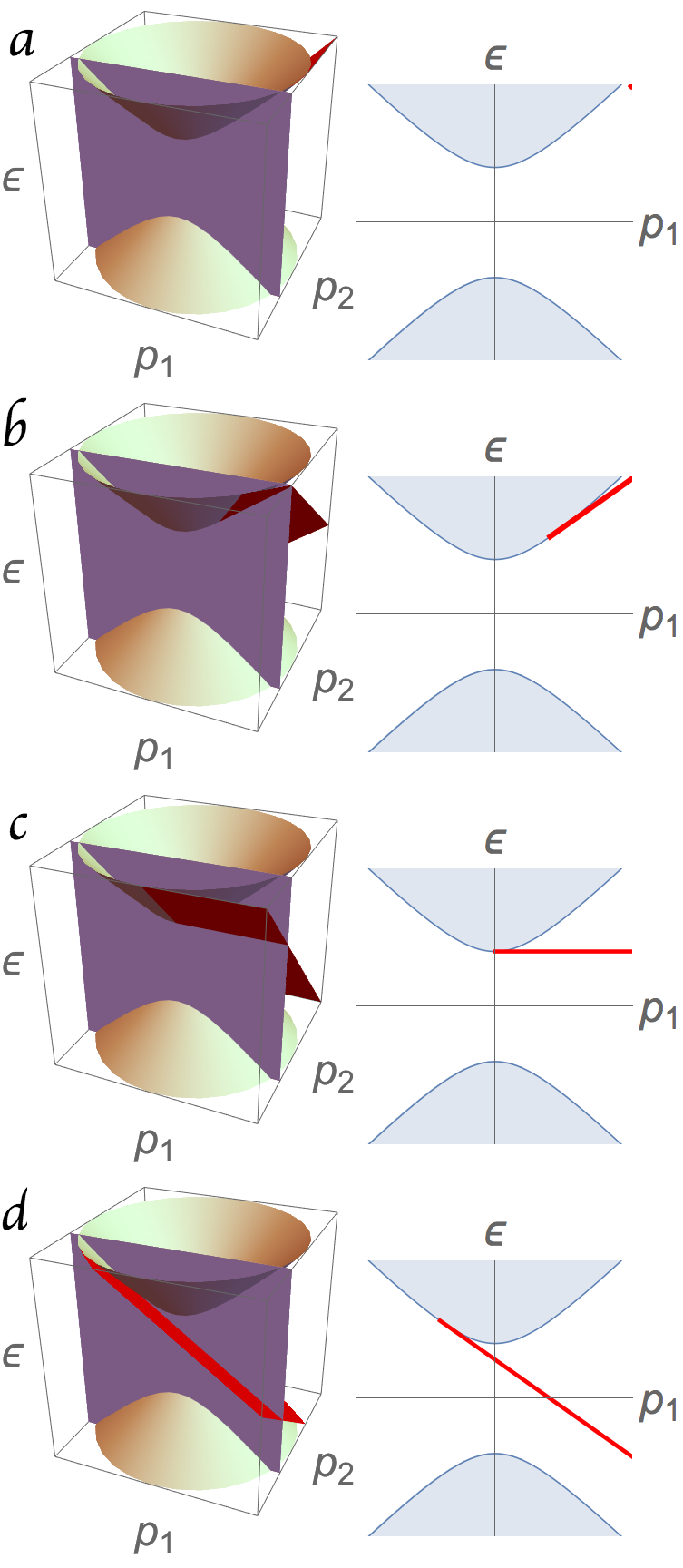}	 
	 \hspace{10mm}
%        \end{center}
%\end{figure}	  
%\begin{figure}[h!]  \label{Dispersion''}
%	\begin{center}
%	 \includegraphics[scale=0.25]{2nd_half.png}
	 \includegraphics[scale=0.25]{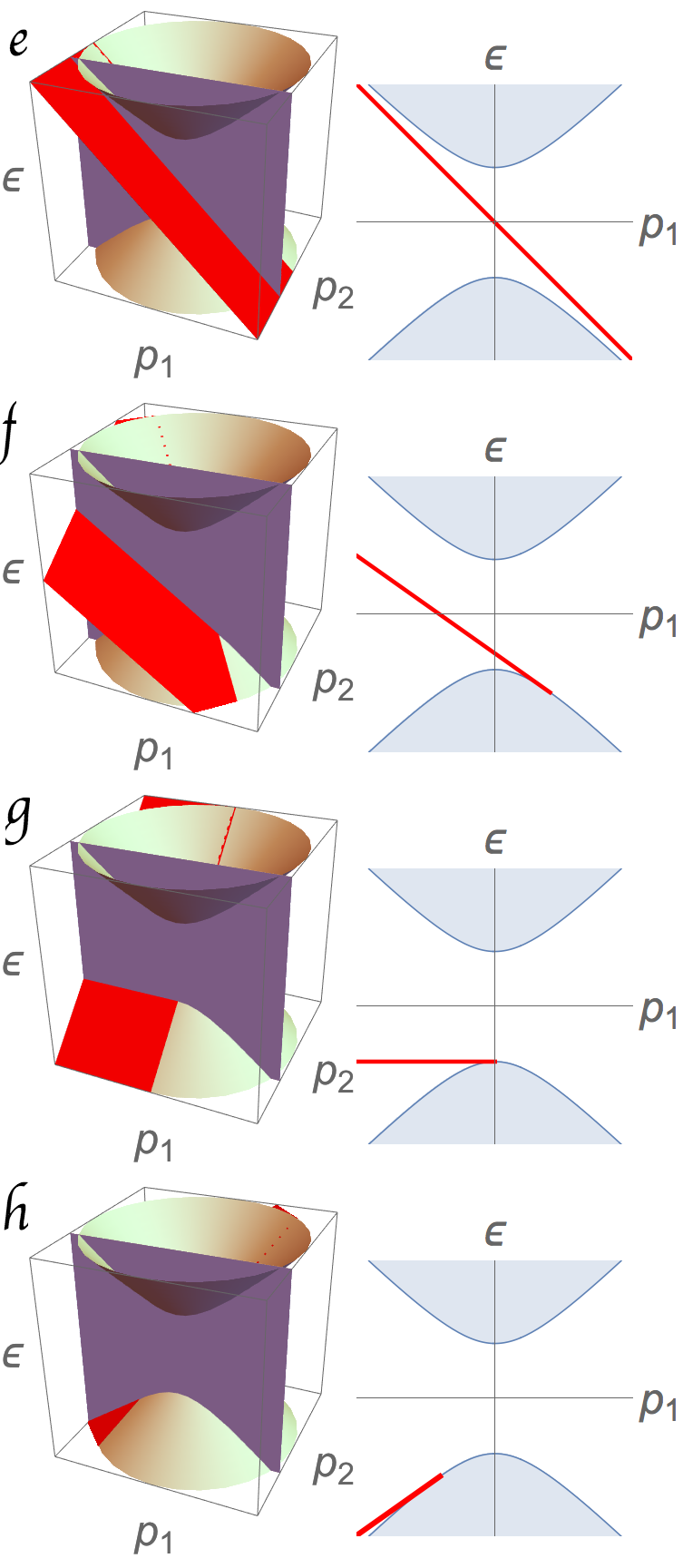}	 
	 \caption{Taking a constant $p_2$ cross-section in the dispersion relation of the 3D case, 
	 we obtain dispersion relation of the 2D topological material, both the gapped bulk states and the edges states. 
	 The red straight half-lines are the edge dispersions of the 2D topological insulator.
	 For each set of figures, the right figure is the cross section while the left figure is the original view of the 3D Weyl semimetals. Figure $(a)$ to $(h)$ have the boundary parameter  
	 $2\theta_+=\pi, \frac{3}{4}\pi, \pi/2, \pi/4,0,  -\pi/4, -\pi/2, -\frac{3}{4}\pi$, respectively.}
	 \label{Dispersion'}
         \end{center}
\end{figure*}

Note that the edge modes exist only in a limited region of the momentum space, since we 
need to require $\alpha>0$. The linear inequality $\alpha>0$ specifies a half of the momentum space,
only in which the dispersion exists, see Fig.~\ref{Dispersion1}. 

In the limit $\alpha = 0$, that is, on the line $p_1\sin{2\theta_+}-p_2\cos{2\theta_+}=0$ in the momentum
space, the edge mode approaches a non-normalizable mode, which is a constant wave function in the $x^3$ space.
It corresponds to $p_3=0$ bulk mode, whose dispersion is $\epsilon = \pm \sqrt{p_1^2+p_2^2}$. In fact, the
edge dispersion (\ref{EDR}) is identical to that under the condition $\alpha=0$. Therefore we have a consistent picture
for any value of $\theta_+$: when the edge mode approaches a non-normalizable state in the momentum space,
it is consistently and continuously absorbed into the bulk modes. In Fig.~\ref{Dispersion1}, we find explicitly that the edge dispersion
surface has its boundary on the bulk dispersion surface.

We would like to make one comment about how we could modify the range of $\alpha$ in different setups. 
If we introduce two boundaries
which are parallel to each other, then the condition of the positivity of $\alpha$ does not apply, as
the wave functions are normalizable even for a negative value of $\alpha$. We demonstrate the
calculation in App.~\ref{sec:2b}.

In summary, we find that the dispersion of the edge state is attached to the bulk Weyl cone
in such a way that (i) the edge dispersion is tangential to the Weyl cone, and (ii) the edge dispersion ends at the touching line on the Weyl cone.

%%%%%%%%%%%%%%%%%%%%%%%%
\subsection{Reduction to 2D}\label{sec:2D_red}
%%%%%%%%%%%%%%%%%%%%%%%%

It is important that the analysis given above can be consistently translated to
2D topological insulator of class A. It is just a dimensional reduction of 
the previous Hamiltonian from 3D to 2D, by a replacement of one of the momenta --
$p_2$ with a constant mass parameter $m$. 
This means that we can study most generic boundary condition of the class A topological
insulator in the continuum limit, and its consequence in the edge dispersions.

By the dimensional reduction, the Hamiltonian of the 2D gapped fermion is given as
\begin{align}
	\mathcal{H}=p_1\sigma_1+m\sigma_2+p_3\sigma_3.
\end{align}
The analysis of the boundary condition we had before for the Weyl semimetals
does not change, %depend on whether $p_2$ is a constant or not, 
since it is just a renaming of $p_2$.
So it is identical to our previous \eqref{bc'}:
\begin{align}
	\left(\begin{array}{cc}1 & e^{-2i\theta_+} 
	\end{array}\right)
	\psi\Big|_{x^3=0}=0.
\end{align}
The dispersion relation $\epsilon$
and the inverse decay width $\alpha$ are given simply by a replacement of $p_2$ with $m$:
\begin{align}
	\left(\begin{array}{c}\epsilon \\\alpha\end{array}\right)
	=
	-\left(\begin{array}{cc}\cos{2\theta_+} & \sin{2\theta_+} \\ -\sin{2\theta_+} & \cos{2\theta_+}\end{array}\right)
	\left(\begin{array}{c}p_1  \\m\end{array}\right).
\end{align}
The same is applied for the edge mode wave function:
\begin{align}
	\psi(x^3)=\sqrt{\alpha}~\text{exp}(-\alpha x^3)
	\left(\begin{array}{c} e^{-2i\theta_+} \\ -1
	\end{array}\right).
\end{align}
Fig.~\ref{Dispersion'} shows the bulk and the edge dispersions for various choices of the boundary parameter $\theta_+$.
Since our procedure is just replacing the momentum $p_2$ by a constant $m$, it amounts to choosing a plane of constant $p_2$ in the 3D Weyl semimetal dispersion given in Fig.~\ref{Dispersion1}. 
Taking a cross-section, we find that the 3D Weyl dispersion and the edge dispersion reduce to 
dispersions of the  gapped bulk and the linear edge modes in 2D.

It is interesting that the rotation $2\theta_+$ in the $(p_1,p_2)$ plane for the 3D Weyl semimetals
can inherits its nature in the 2D topological gapped system in an nontrivial manner. The form
of the edge dispersion, as a function of $p_1$, looks quite nontrivial in Fig.~\ref{Dispersion'}. For some special choice of
the value $\theta_+ = \pi/2$, the edge dispersion eventually disappear. For some other values of
$\theta_+$, the edge dispersion becomes a flat band.

By taking a massless limit $m=0$ for the bulk system, %(such as graphene),
the edge dispersion is simply given by
\begin{align}
\epsilon = -p_1 \cos (2\theta_+) \, .
\end{align}
The existence condition of the edge state is $p_1 \sin 2\theta_+ (=\alpha) > 0$. 
So the edge dispersion, which is a half line, emanates linearly from the Dirac point of 
the graphene by the slope $\cos 2\theta_+$, 
where the parameter $\theta_+$ can range $0\leq  \theta_+ < \pi$.

%%%%%%%%%%%%%%%%%%%%%%%%%%%%%%%%%%%%%

\section{Lattice models}\label{sec:lattice}

%%%%%%%%%%%%%%%%%%%%%%%%%%%%%%%%%%%%%

The effective model study shown above exhibits an interesting behavior of the edge state depending on the boundary condition.
Let us then show how such an argument on the boundary condition is realized in lattice models 
with tight-binding Hamiltonians.

\subsection{Boundary condition for discretized model}

In the effective continuum theory the boundary condition requires some conditions due to self-conjugacy of the Hamiltonian.
Following this argument, we consider the boundary condition with the discretized lattice model.

First of all, we cannot directly apply the continuum theory argument to the lattice model because this argument relies on the integral by parts:
We need to replace the differential operator with a difference operator which does not satisfy the Leibniz rule.
We have to be careful about dealing with the boundary of the discrete lattice system.

To demonstrate how the self-conjugacy characterizes the boundary condition, we consider a discrete model defined on a finite one-dimensional lattice labeled by $n = 1, \ldots, N$.
The self-conjugate operator we consider here is $\mathcal{H} = - i \sigma \nabla$ where $\sigma$ is a Hermitian matrix to be taken as a Pauli matrix, and the difference operator is defined
\begin{align}
 \nabla \psi_n & = \psi_{n+1} - \psi_n
 \, , \\
 \nabla^\dag \psi_n & = \psi_{n-1} - \psi_n
 \, .
\end{align}
This difference operator reduces to the differential operator in the continuum limit, so that the operator becomes the standard Dirac Hamiltonian $\mathcal{H} \to -i \sigma \partial_x$ in the limit.
Since they are related to each other, $i\nabla^\dag \psi_{n+1} = -i\nabla \psi_n$, this is locally self-conjugate.
However, as discussed before, we need to take care of the boundary:
The discrete Dirac Hamiltonian is self-conjugate up to the boundary term
\begin{align}
 \sum_{n=1}^N \psi^\dag_n \left( -i \sigma \nabla \psi_n \right) 
 & =
 \sum_{n=1}^N \left( i \sigma \nabla^\dag \psi_n \right)^\dag \psi_n
 \nonumber \\
 & \quad
 + \psi^\dag_0 (i\sigma) \psi_1
 - \psi^\dag_N (i\sigma) \psi_{N+1}
\end{align}
where we introduced auxiliary fields $\psi_0$ and $\psi_{N+1}$.
The second line shows the surface term in this case, and the self-conjugacy of the Hamiltonian requires that this part should vanish
\begin{align}
 \psi^\dag_0 (i\sigma) \psi_1
 - \psi^\dag_N (i\sigma) \psi_{N+1} = 0
 \, .
\end{align}
We have two possibilities to solve this condition.
The first is the periodic boundary condition $\psi_n = \psi_{n+N}$ for $\forall n \in \{1, \ldots, N\}$.
Then these two terms cancel each other.
The second is the situation that the both two terms vanish independently, which corresponds to the open boundary condition.

Let us focus on the first term $\psi^\dag_0 (i\sigma) \psi_1$, and apply the boundary condition which is analogous to that considered in continuum theory
\begin{align}
 \left( M + 1 \right) \psi\Big|_{n=0,1} = 0
 \, .
\end{align}
We remark that this boundary condition is assigned to both $\psi_{n=0}$ and $\psi_{n=1}$, although the former one is just an auxiliary field.
If this matrix $M$ satisfies
\begin{align}
 M^\dag \sigma + \sigma M = 0
 \, ,
 \label{eq:M_cond_lat}
\end{align}
the first term vanishes due to the same argument given in the continuum theory shown in Sec.~\ref{sec:bc_cont}.
We also apply a similar condition to the opposite boundary $n = N , N+1$ with a matrix $M'$ which is not necessarily the same as $M$ as long as satisfying the condition \eqref{eq:M_cond_lat}.

\begin{figure}[t]
 \begin{center}
  \begin{tikzpicture}[thick]

   \node at (0,0) {\includegraphics[width=17em]{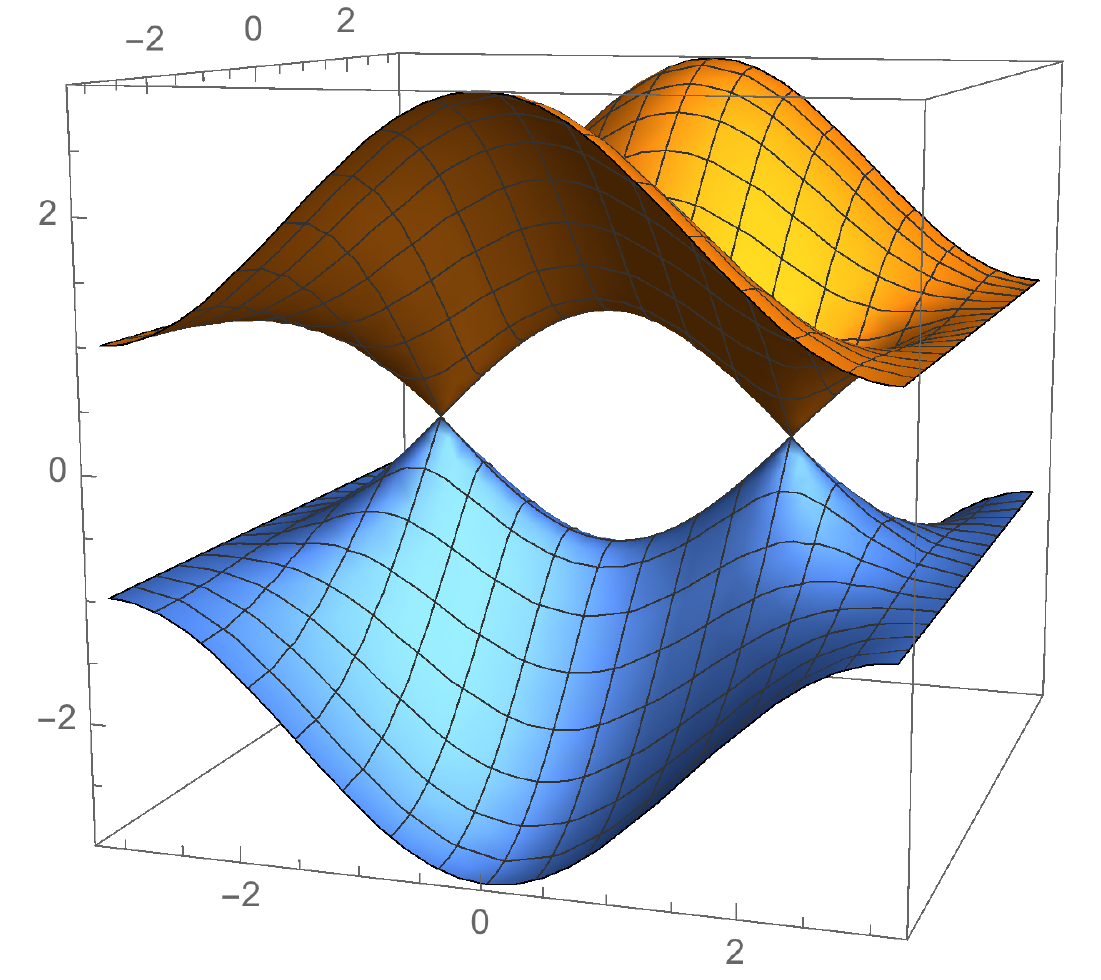}};

    \begin{scope}[shift={(-4.5,1)}]

   \draw[->] (0,0) %(-2.7,-2.1) node (O) {}
     -- ++(-6.8:1) node [right] {$p_1$};
   \draw[->] (0,0) -- ++(90:1.2) node [above] {$\epsilon$};
   \draw[->] (0,0) -- ++(34:1.2) node [above right] {$p_2$};
   
    \end{scope}
   
  \end{tikzpicture}
 \end{center}
 \caption{The energy band spectrum of the bulk Hamiltonian \eqref{eq:3D_Bloch_Ham} with $p_3=0$ and $c=1$. The Weyl points are at $(p_1,p_2)=(\pm \pi/2,0)$.}
 \label{fig:3D_bulk_band}
\end{figure}
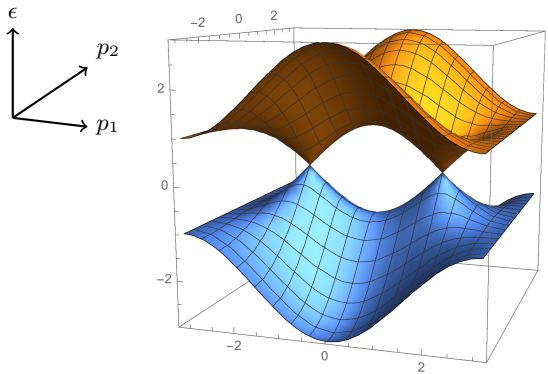

\subsection{3D Weyl semimetal model}

We then incorporate the boundary condition to the Weyl semimetal model on a lattice.
We consider the Hamiltonian defined on a 3D lattice,
\begin{align}
 H = \sum_{n} \psi_{n}^\dag \mathcal{H} \psi_n
\end{align}
where $n$ is a three-dimensional vector $n = (n_1, n_2, n_3) \in \mathbb{Z}^3$, and the operator is given by
\begin{align}
 \mathcal{H} & =
 \frac{1}{2} \sigma_1 \left( \nabla_1 + \nabla_1^\dag - \nabla_2 - \nabla_2^\dag + 2c \right) 
 \nonumber \\
 & \qquad
 - \frac{i}{2} \sigma_2 \left( \nabla_2 - \nabla_2^\dag \right)
 - \frac{i}{2} \sigma_3 \left( \nabla_3 - \nabla_3^\dag \right) 
 \, .
\end{align}
In the Fourier basis, the Bloch Hamiltonian is obtained as
\begin{align}
 \mathcal{H}(p) & =
 \sigma_1 (\cos p_1 - \cos p_2 + c ) 
 + \sigma_2 \sin p_2 + \sigma_3 \sin p_3
 \, ,
 \label{eq:3D_Bloch_Ham}
\end{align}
which exhibits Weyl points at
\begin{align}
 (p_1, p_2, p_3) & =
 \begin{cases}
  (\cos^{-1}(1-c), 0, 0 \ \& \ \pi) & (0 \le c \le 2) \\
  (\cos^{-1}(-1-c), \pi, 0 \ \& \ \pi) & (-2 \le c \le 0) \\
  \text{n/a} & (|c| > 2)
 \end{cases}
 \, .
\end{align}
The parameter $c$ tunes the Weyl point positions.
The energy band spectrum is drawn in Fig.~\ref{fig:3D_bulk_band} with $p_3 = 0$ and $c=1$.
We see two Weyl points at $(p_1,p_2) = (\pm \pi/2, 0)$ at this section.

We introduce the boundary to this model.
Suppose the lattice is defined on the region $n_3 \ge 1$, and impose the boundary condition
\begin{align}
 \left( M + 1 \right) \psi\Big|_{n_3=1} = 0
 \label{eq:bc_3d_lat}
\end{align}
where the matrix $M$ satisfies
\begin{align}
 M^\dag \sigma_3 + \sigma_3 M = 0
 \, .
\end{align}
Then the situation is completely parallel with the continuum theory studied in the previous section.
The matrix $M$ is parametrized by two parameters, $\theta_+$ and $\theta_-$, and the boundary condition is rephrased in terms of these parameters,
\begin{align}
 \begin{pmatrix}
  1 & e^{-2i\theta_+}
 \end{pmatrix}
 \psi \Big|_{n_3=1} & = 0
 \, ,
 \label{eq:bc_lat3D}
\end{align}
which is equivalent to
\begin{align}
 \psi_{n_3 = 1} \propto
 \begin{pmatrix}
  1 \\ - e^{2 i \theta_+}
 \end{pmatrix}
 \, .
 \label{eq:bc_lat3D2} 
\end{align}
Thus it depends only on the parameter $\theta_+$ in the end.

\begin{figure}[t]
 \begin{center}
  \includegraphics[width=10em]{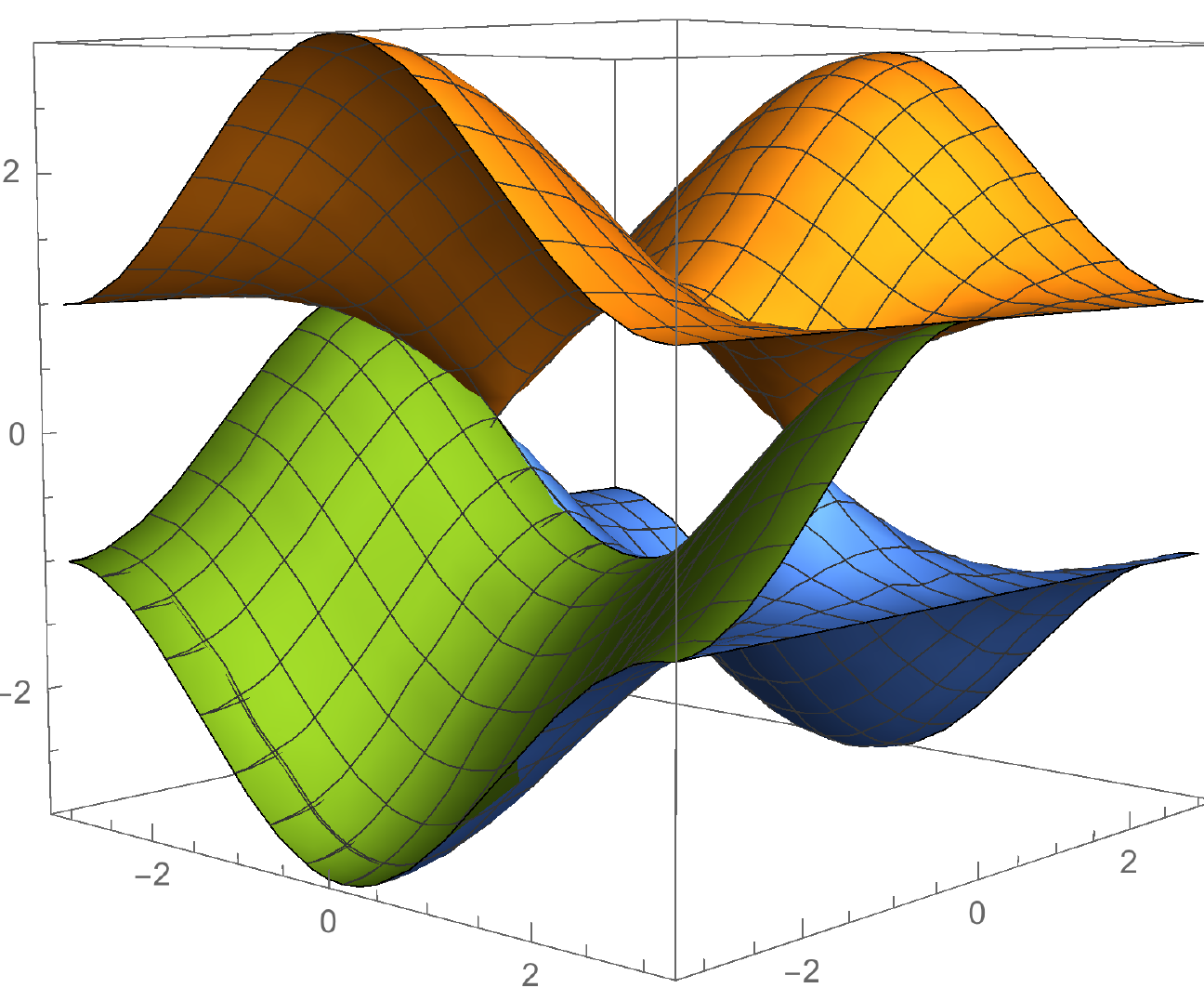} \qquad
  \includegraphics[width=10em]{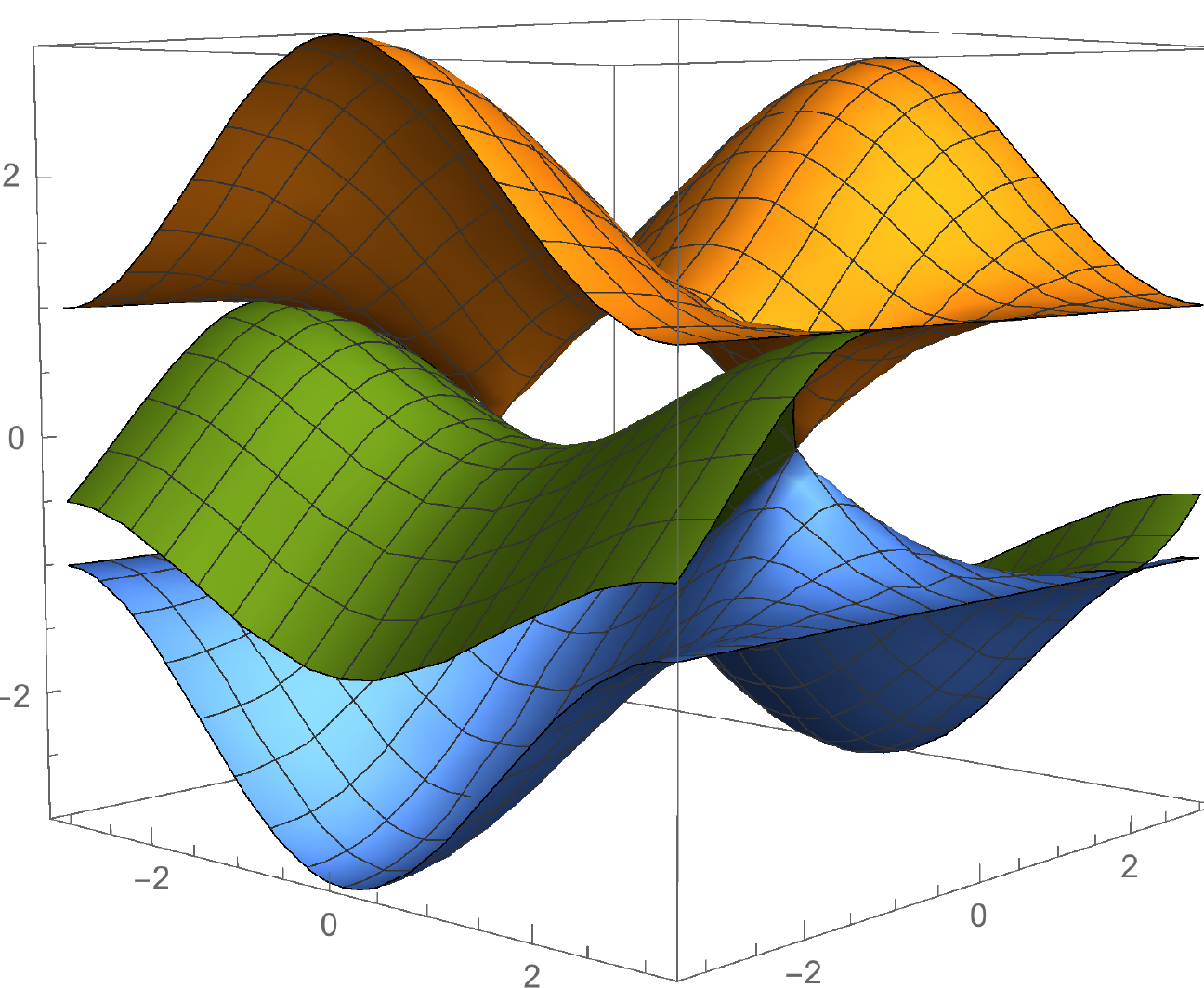} \\[2em]
  \includegraphics[width=10em]{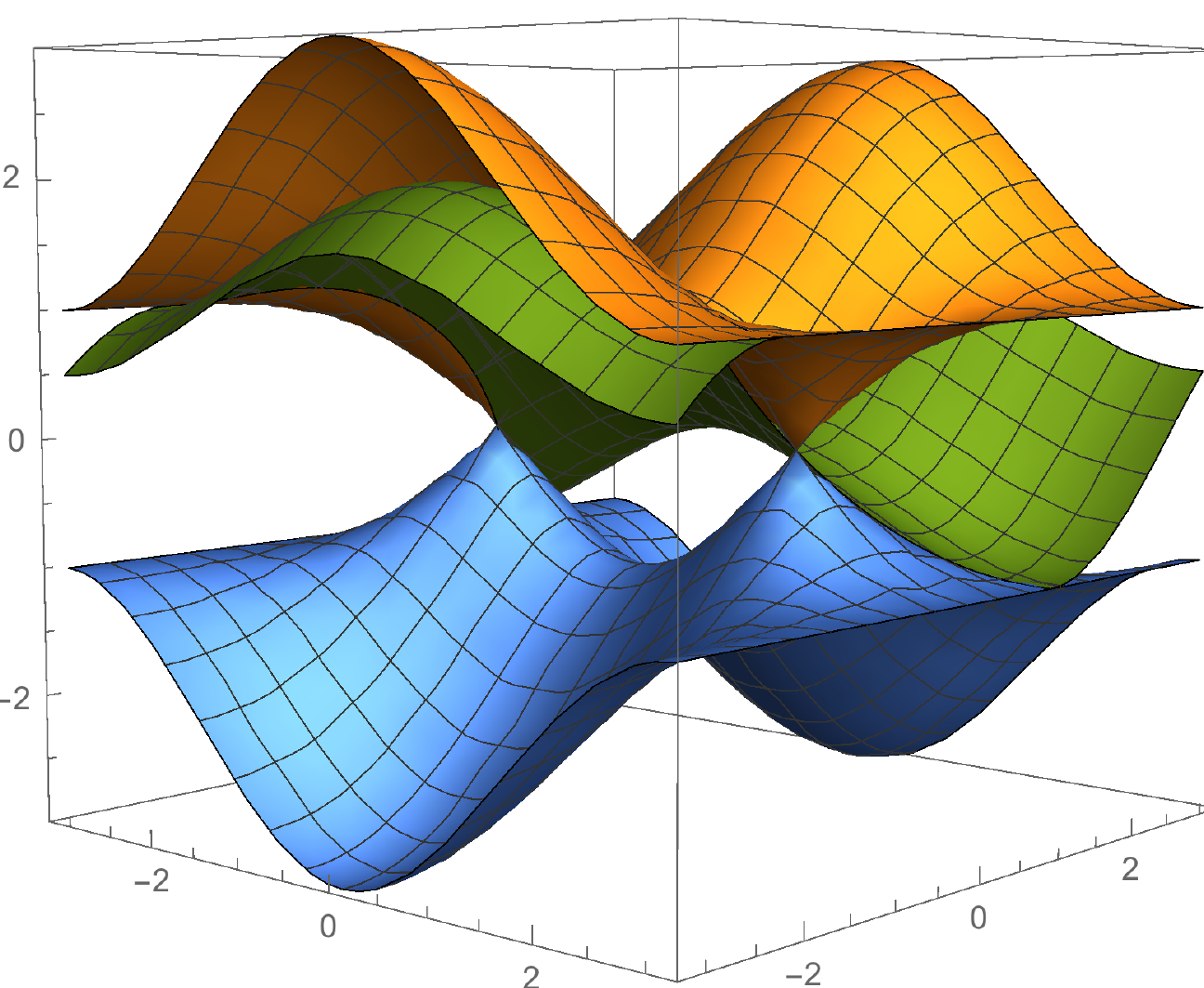} \qquad
  \includegraphics[width=10em]{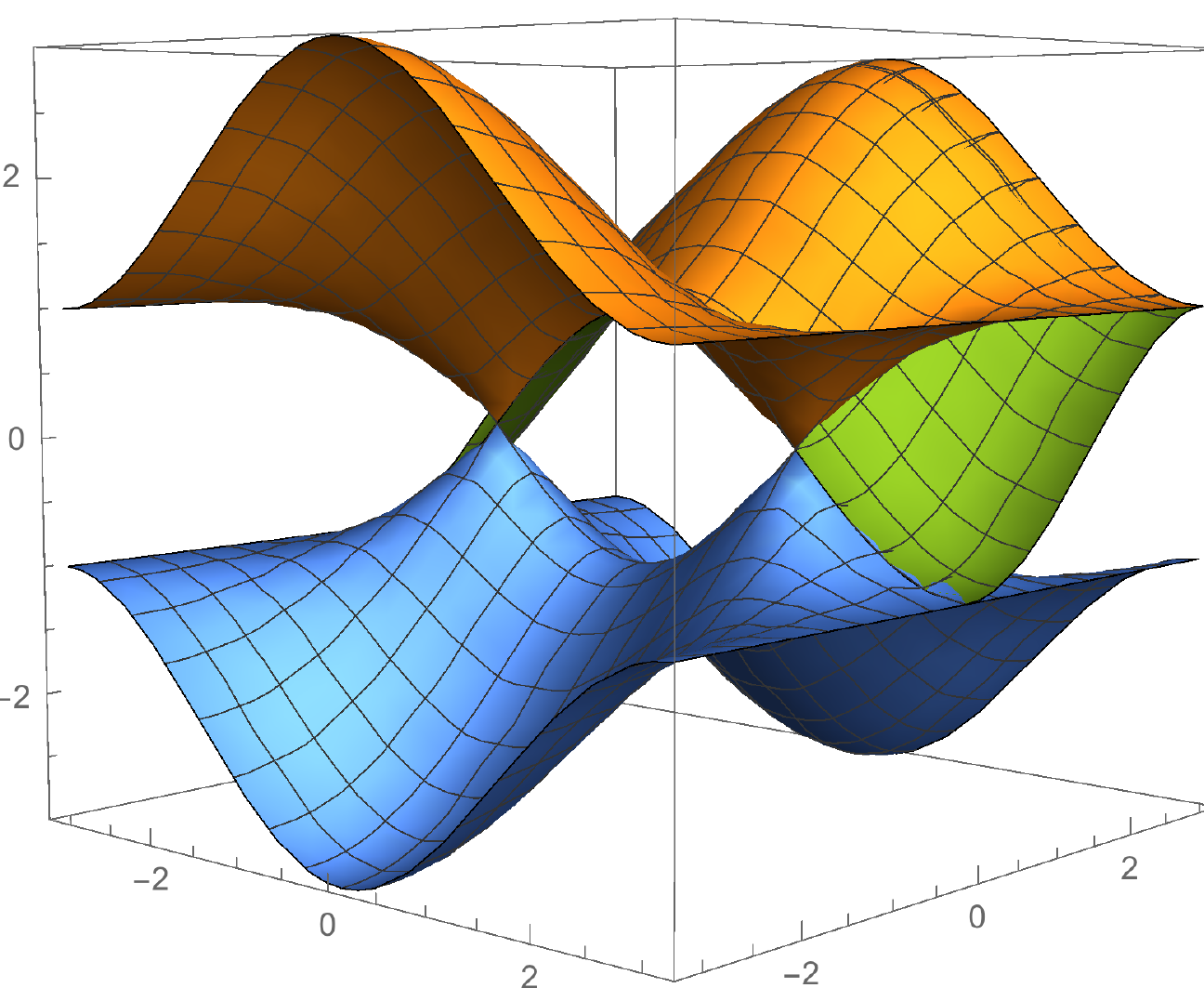}
 \end{center}
 \caption{The dispersion relations of the bulk (orange and blue) and the edge (green) states with respect to the $(p_1,p_2)$-plane (horizontal) for positive $\beta$ and $2\theta_+ = 0, \pi/3, 2\pi/3, \pi$.}
 \label{fig:edge_disp_lat3D}
\end{figure}

We consider the spectrum and wave function behavior of the edge state under the boundary condition \eqref{eq:bc_3d_lat}.
For the eigenvalue equation
\begin{align}
 \mathcal{H} \psi = \epsilon(p) \psi
 \, ,
 \label{eq:eigen_prob_lat}
\end{align}
the Hamiltonian has a matrix form in the partial Fourier basis,
\begin{align}
 \mathcal{H}
% & =
% \sigma_1 \left( \cos p_1 - \cos p_2 + m \right)
% \nonumber \\
% & \qquad
% + \sigma_2 \sin p_2
% - \frac{i}{2} \sigma_3 \left( \nabla_3 - \nabla_3^\dag \right)
% \nonumber \\
 & =
 \begin{pmatrix}
  0 & \Delta(p)^\dag \\
  \Delta(p) & 0
 \end{pmatrix}
 - \frac{i}{2} \sigma_3 \left( \nabla_3 - \nabla_3^\dag \right)
\end{align}
where the off-diagonal element is given by
\begin{align}
 \Delta(p) & =
 \cos p_1 - \cos p_2 + c + i \sin p_2
 \, ,
 \label{eq:Delta}
\end{align}
which behaves as $\Delta(p) \sim p_1 \pm i p_2$ in the vicinity of the Weyl points. 
The sign $\pm$ depends on the chirality of the Weyl points.

\begin{figure}[t]
 \begin{center}
  \begin{tikzpicture}%[thick]

   \node (1) at (0,0) {\includegraphics[width=10em]{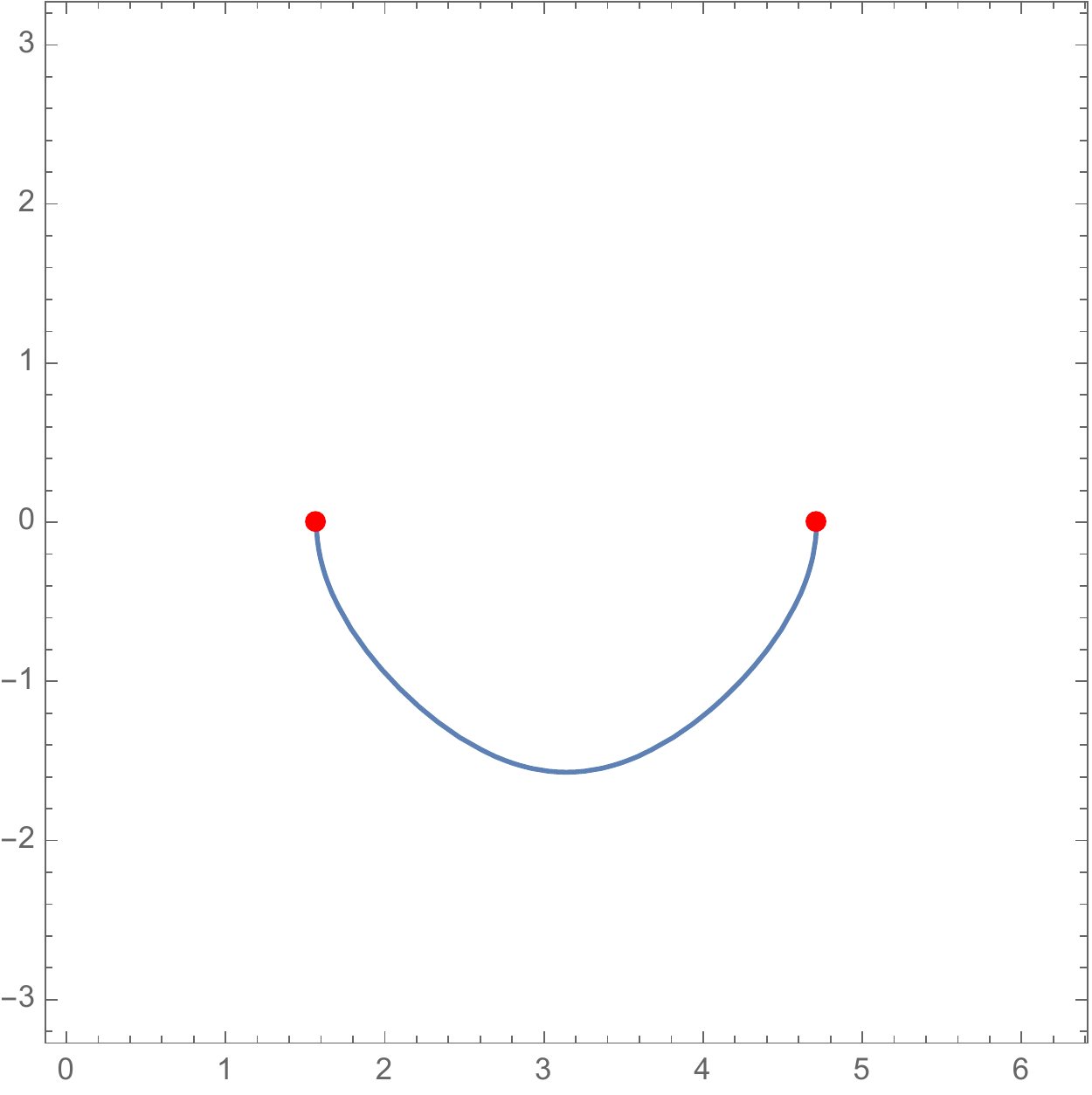}};
   \node (2) at (4,0) {\includegraphics[width=10em]{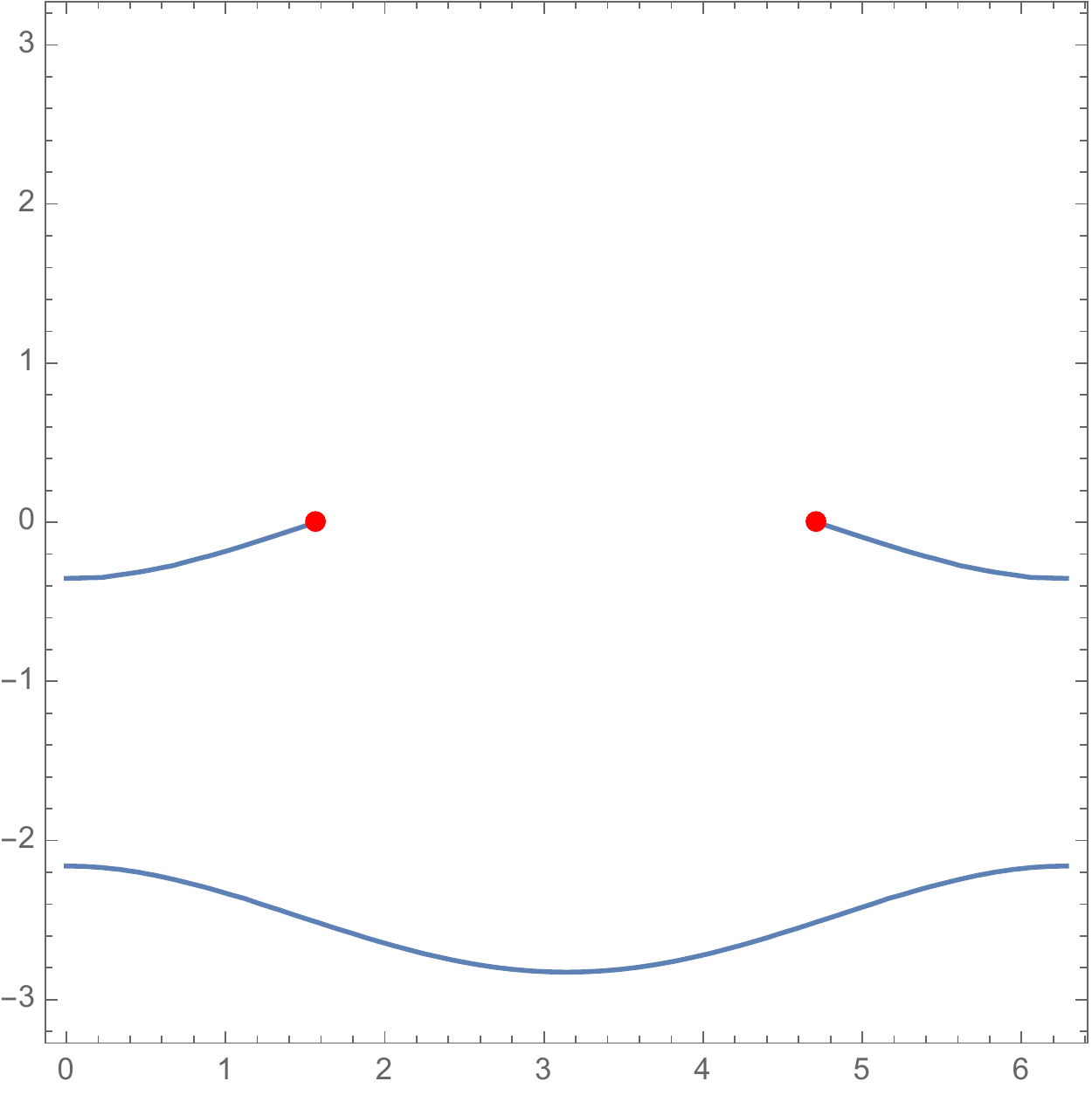}};
   \node (3) at (0,-4) {\includegraphics[width=10em]{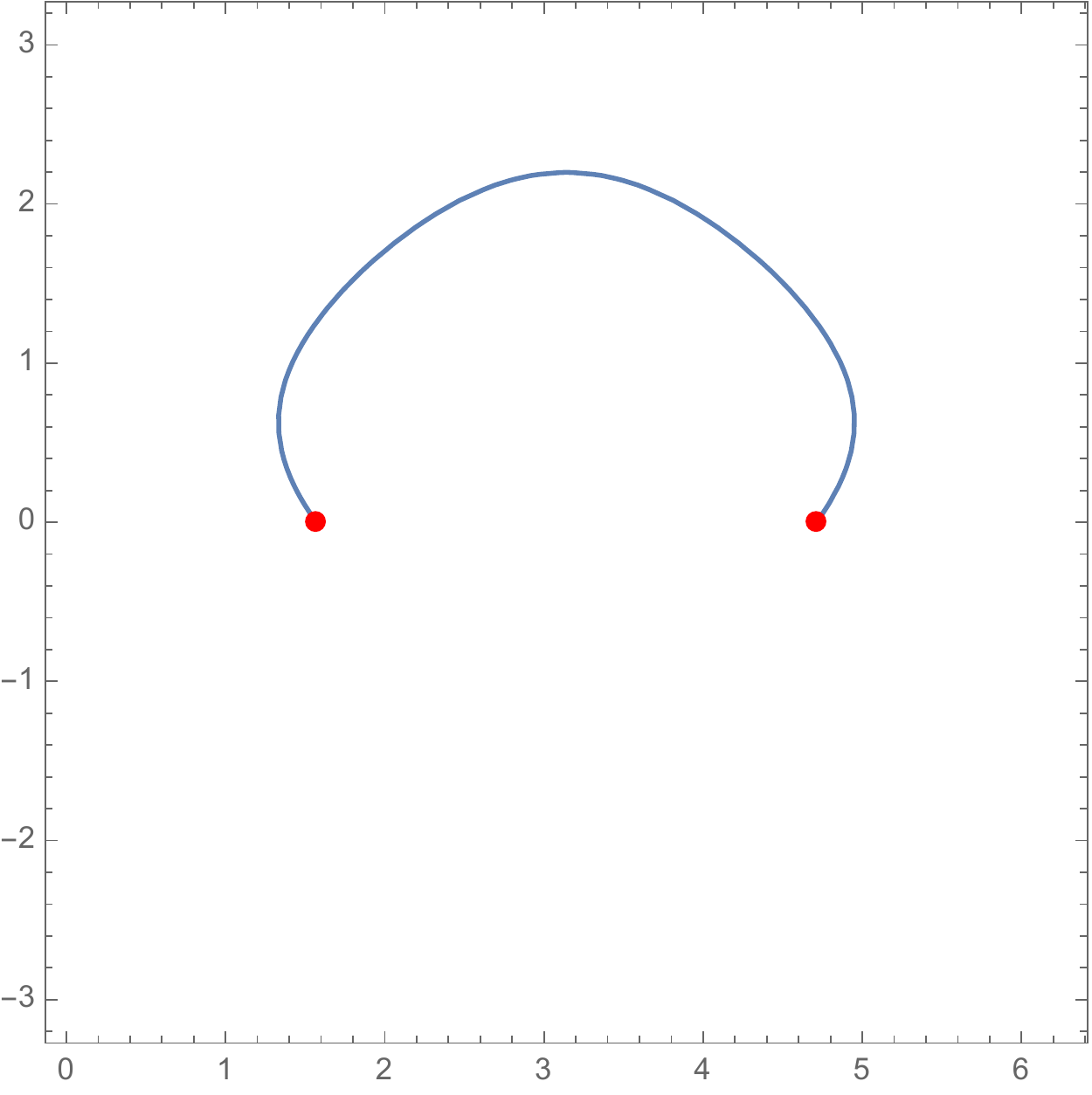}};
   \node (4) at (4,-4) {\includegraphics[width=10em]{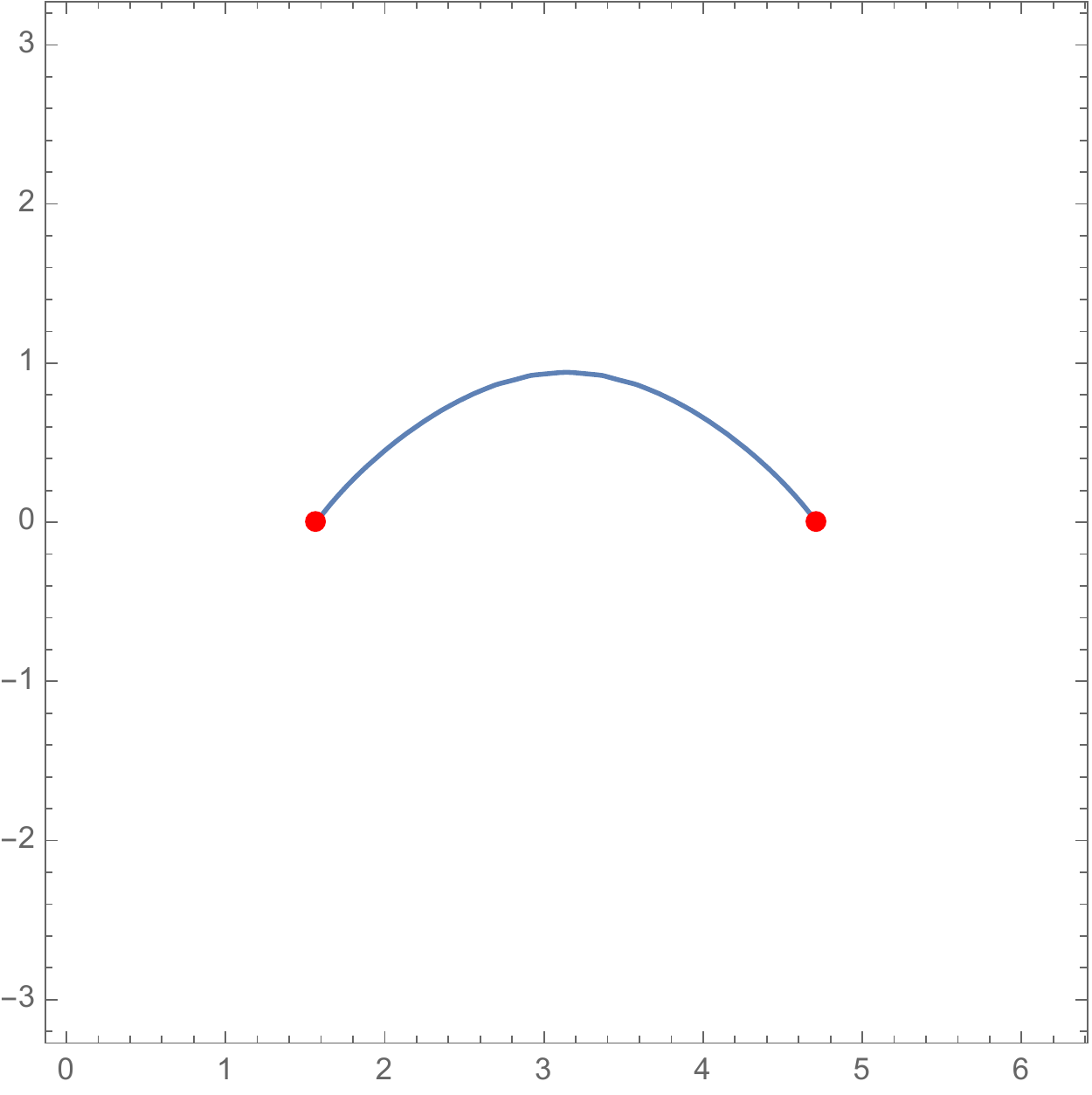}};
   \node (5) at (0,-8) {\includegraphics[width=10em]{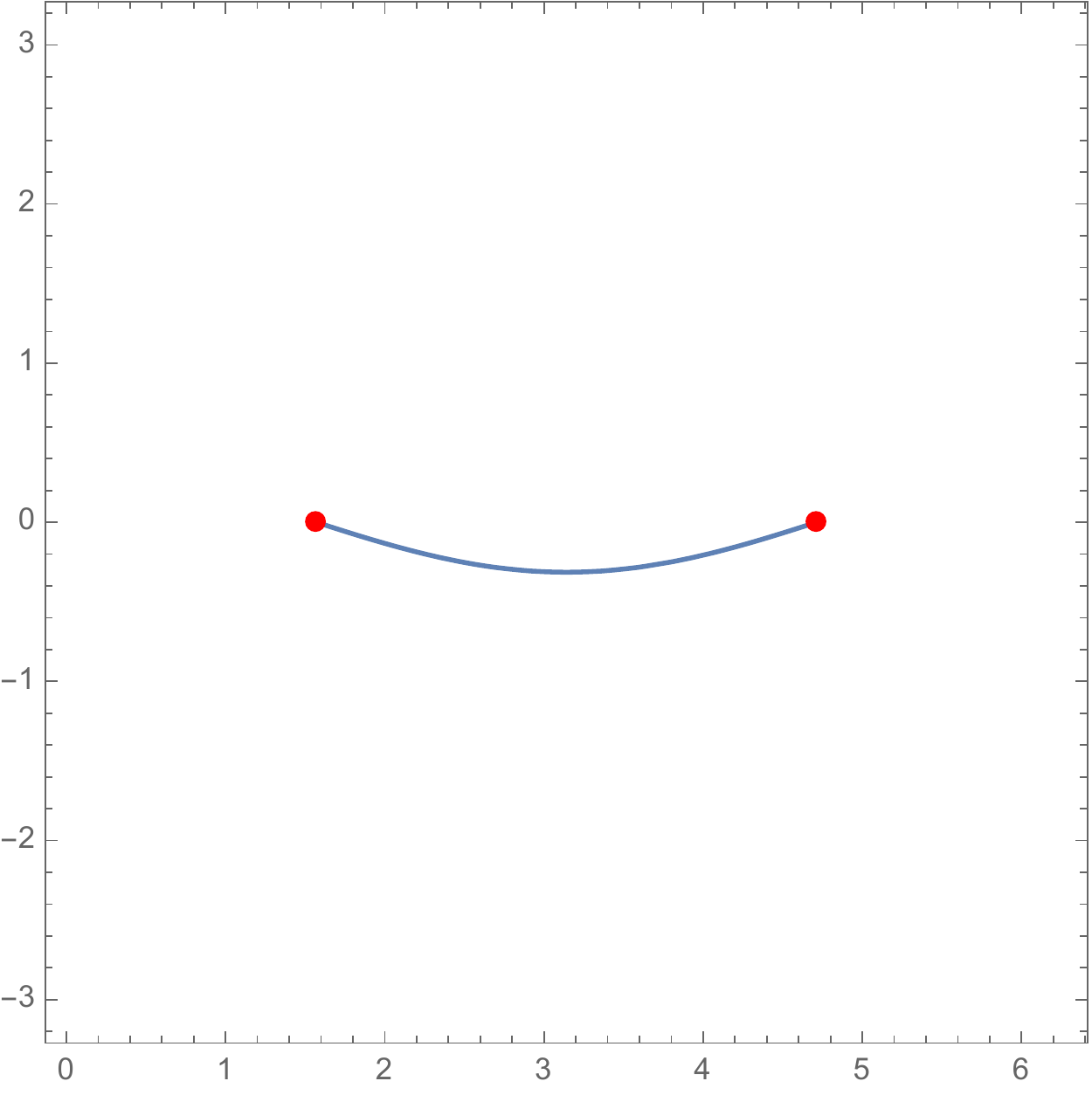}};
   \node (6) at (4,-8) {\includegraphics[width=10em]{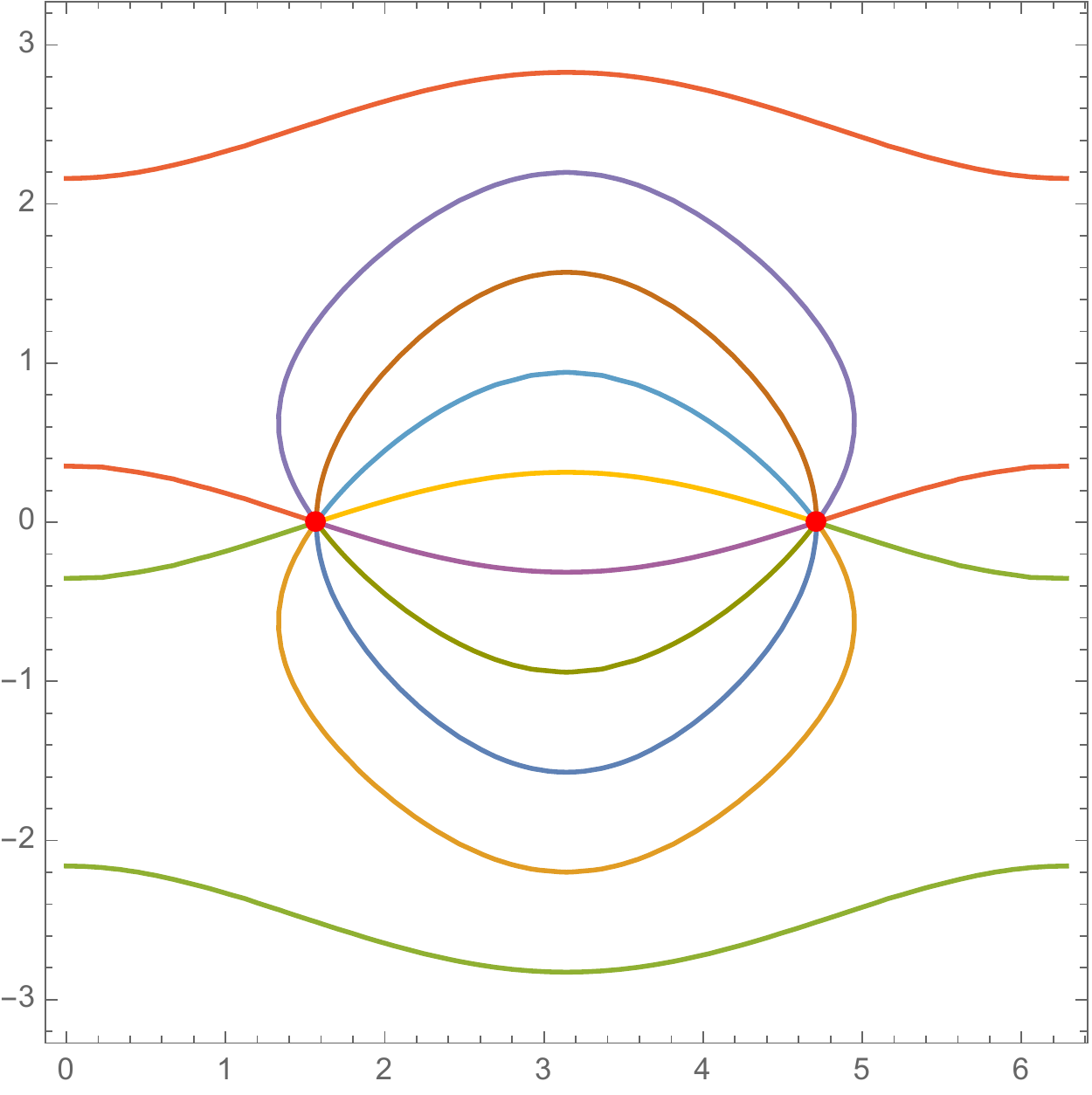}};

   \node at (.8,1.3) {$2\theta_+ = 0$};
   \node at (4.8,1.3) {$2\theta_+ = \frac{2\pi}{5}$};
   \node at (.8,1.3-4) {$2\theta_+ = \frac{4\pi}{5}$};
   \node at (4.8,1.3-4) {$2\theta_+ = \frac{6\pi}{5}$};
   \node at (.8,1.3-8) {$2\theta_+ = \frac{8\pi}{5}$};

   \begin{scope}[shift={(4,-8)}]

    \draw[->] (.8,-.2) arc (-90:250:.3);
    \draw[->] (-.7,-.2) arc (270:-50:.3);

    \node at (1.3,.5) {$\theta_+$};
    \node at (-1.1,.5) {$\theta_+$};
    
   \end{scope}
   
  \end{tikzpicture}

 \end{center}
 \caption{The parameter dependence of the Fermi arc which is the zero energy section $\epsilon(p) = 0$ of the edge state energy spectrum with $2 \theta_+ = 0, 2\pi/5,4\pi/5,6\pi/5,8\pi/5$ for positive $\beta$. The horizontal and vertical axes are for $p_1$ and $p_2$. The red dot shows the bulk Weyl node. The last panel shows Fermi arcs with various values of the parameter $\theta_+ \in [0,\pi)$.}
 \label{fig:Fermi_arc1}
\end{figure}
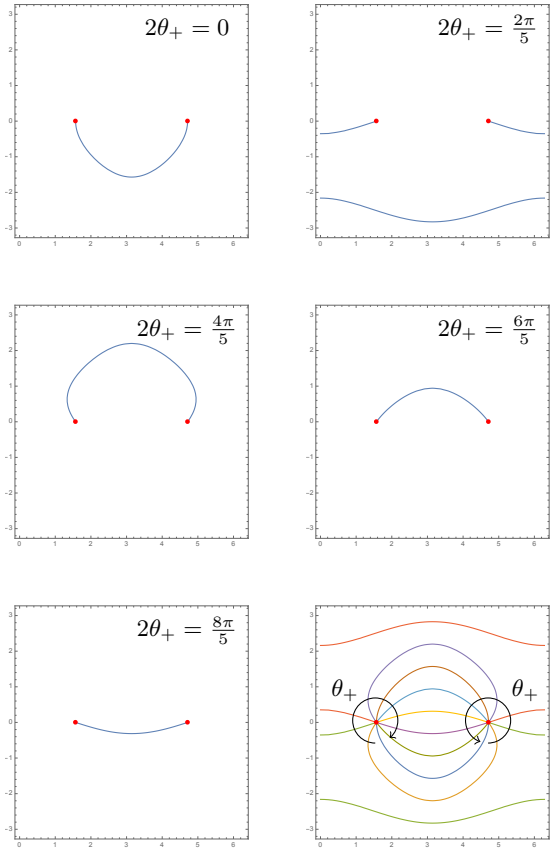

\begin{figure}[t]
 \begin{center}

  \begin{tikzpicture}%[thick]

   \node (1) at (0,0) {\includegraphics[width=10em]{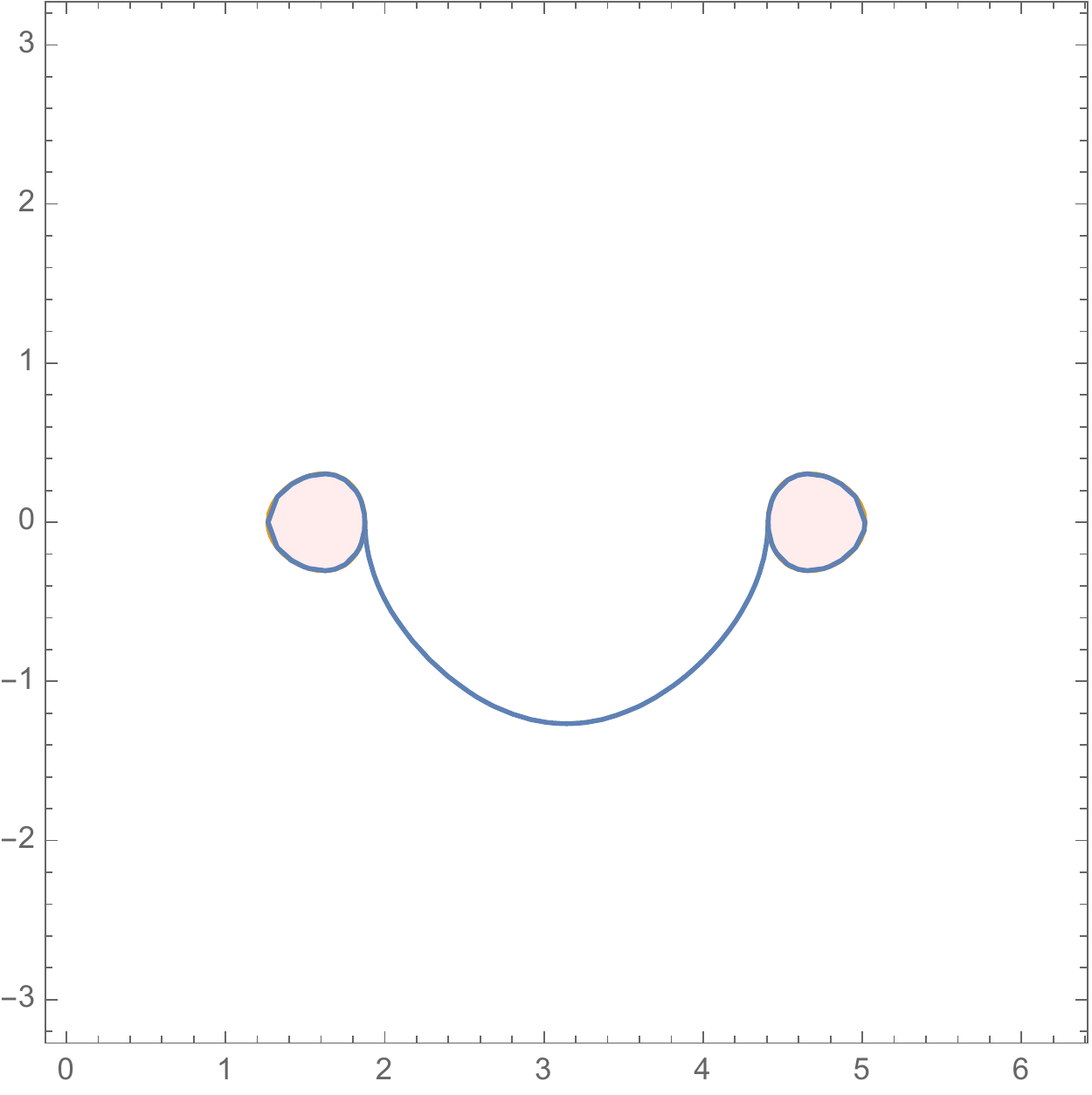}};
   \node (2) at (4,0) {\includegraphics[width=10em]{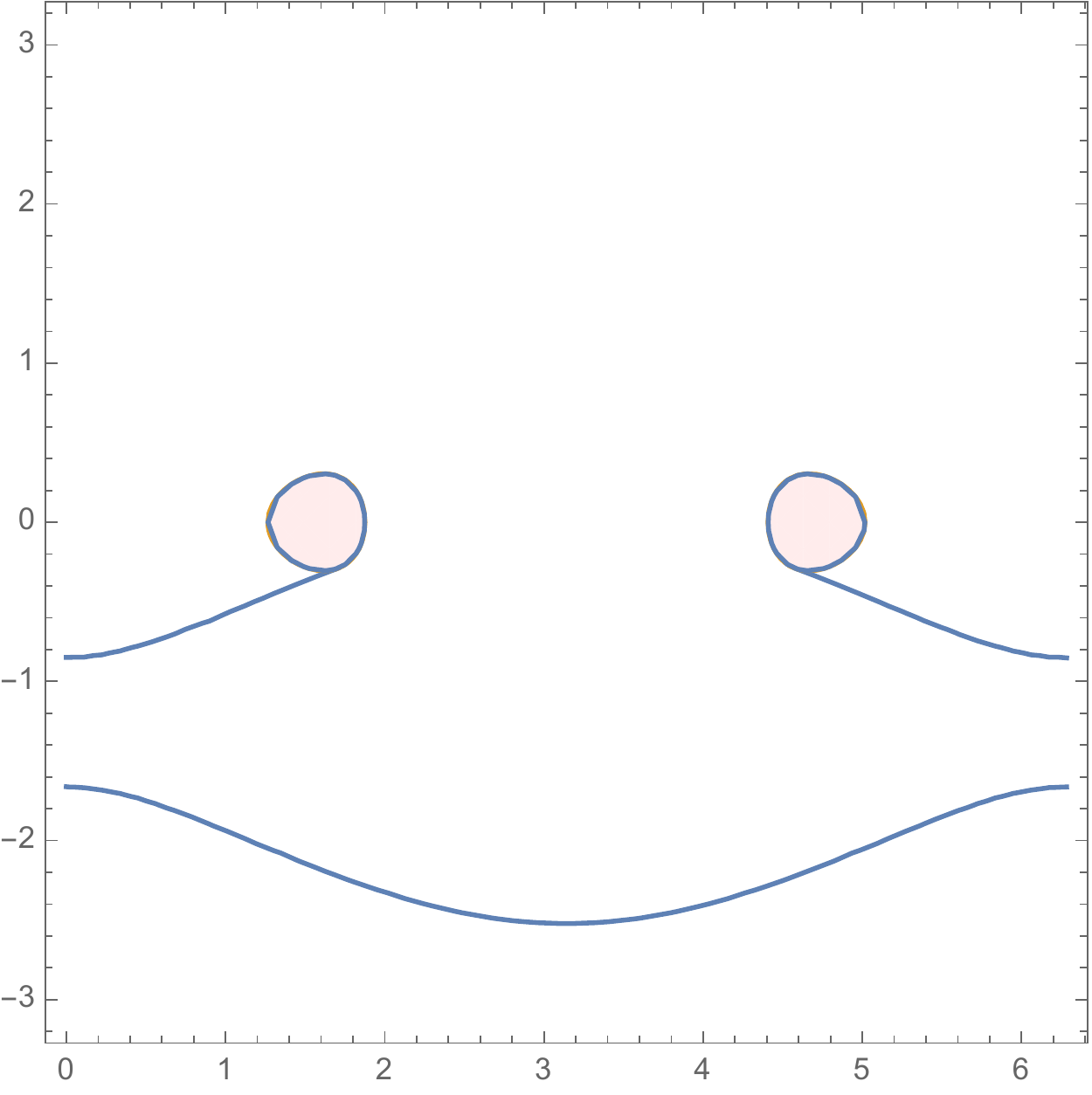}};
   \node (3) at (0,-4) {\includegraphics[width=10em]{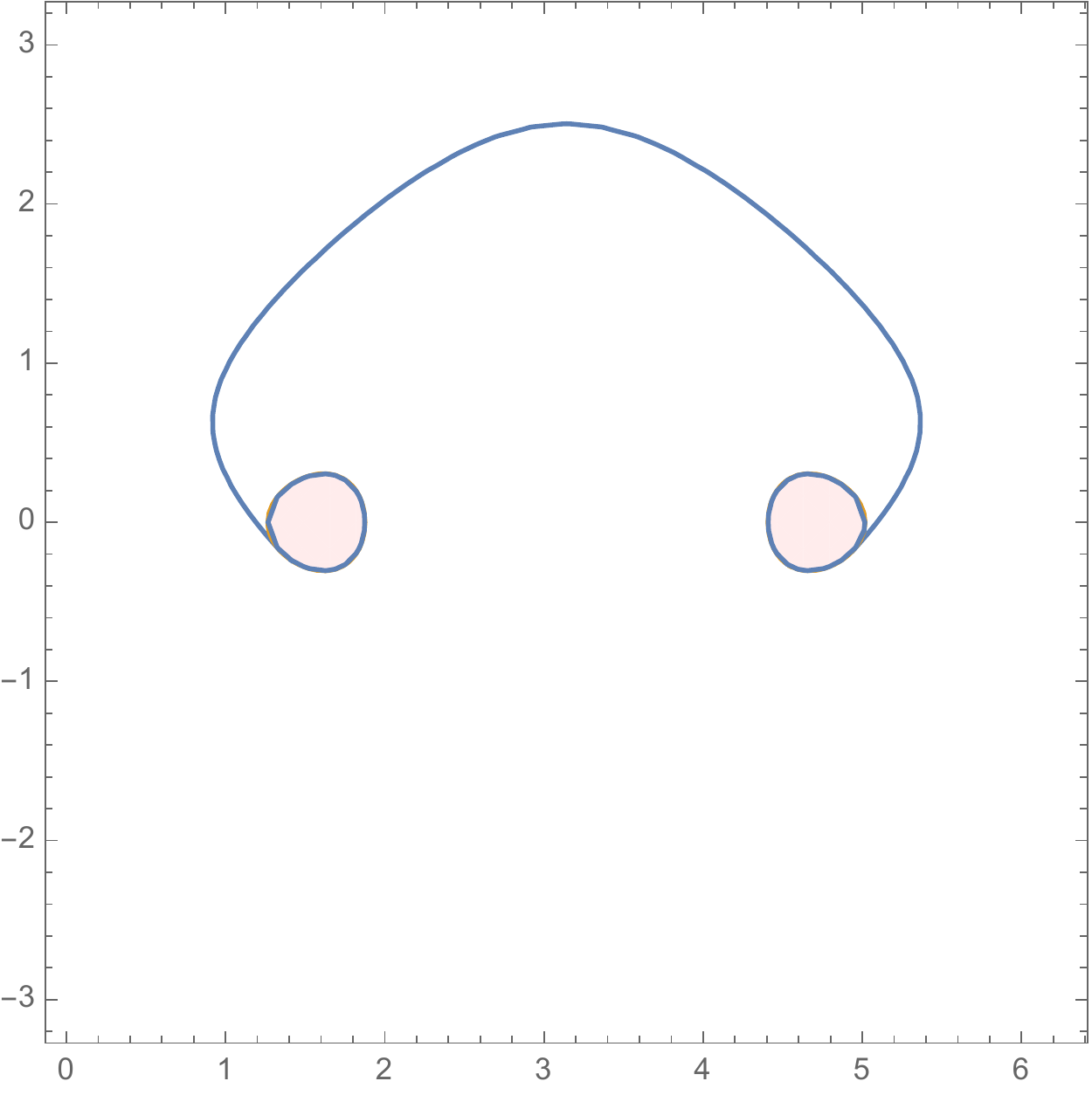}};
   \node (4) at (4,-4) {\includegraphics[width=10em]{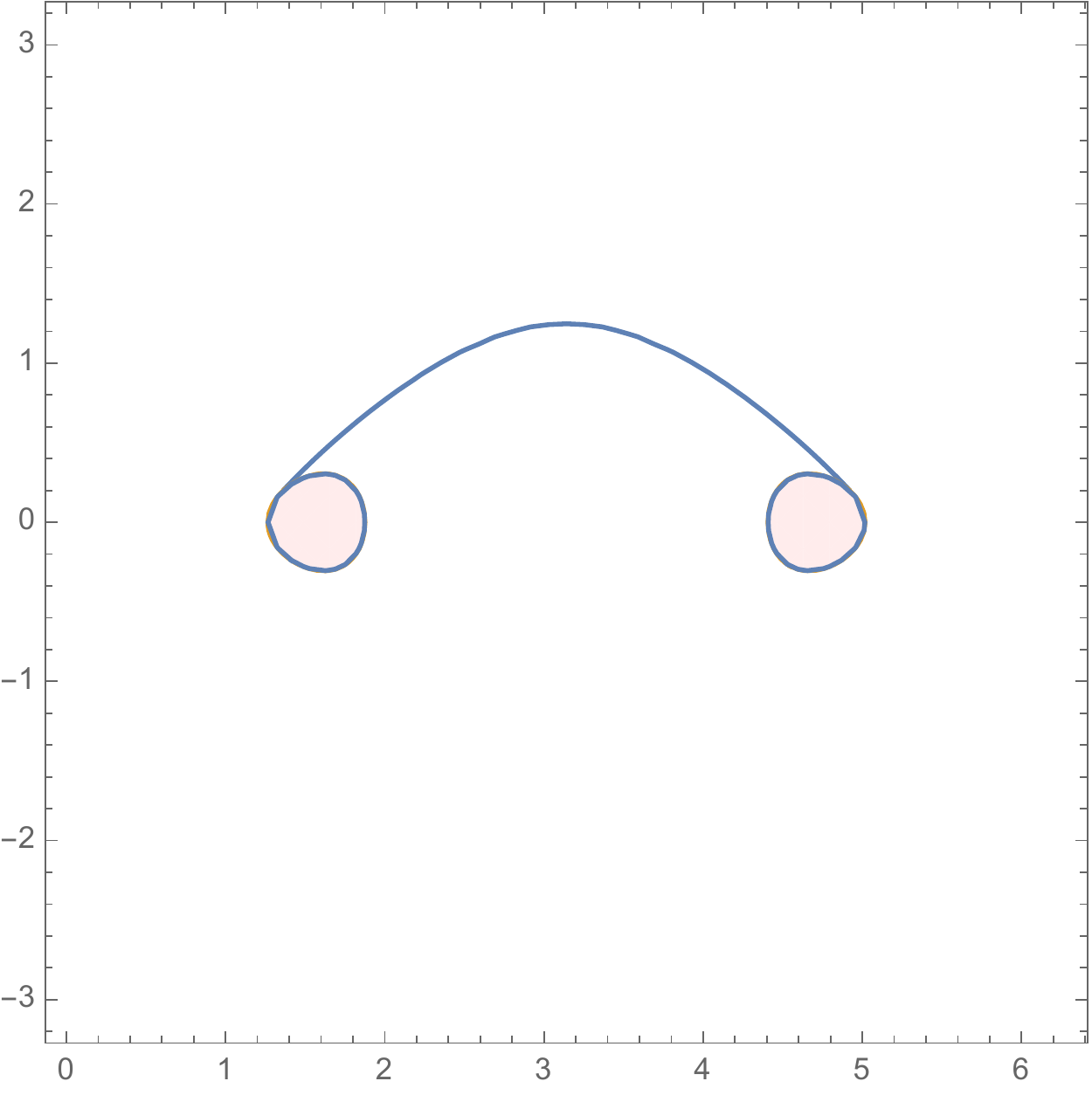}};
   \node (5) at (0,-8) {\includegraphics[width=10em]{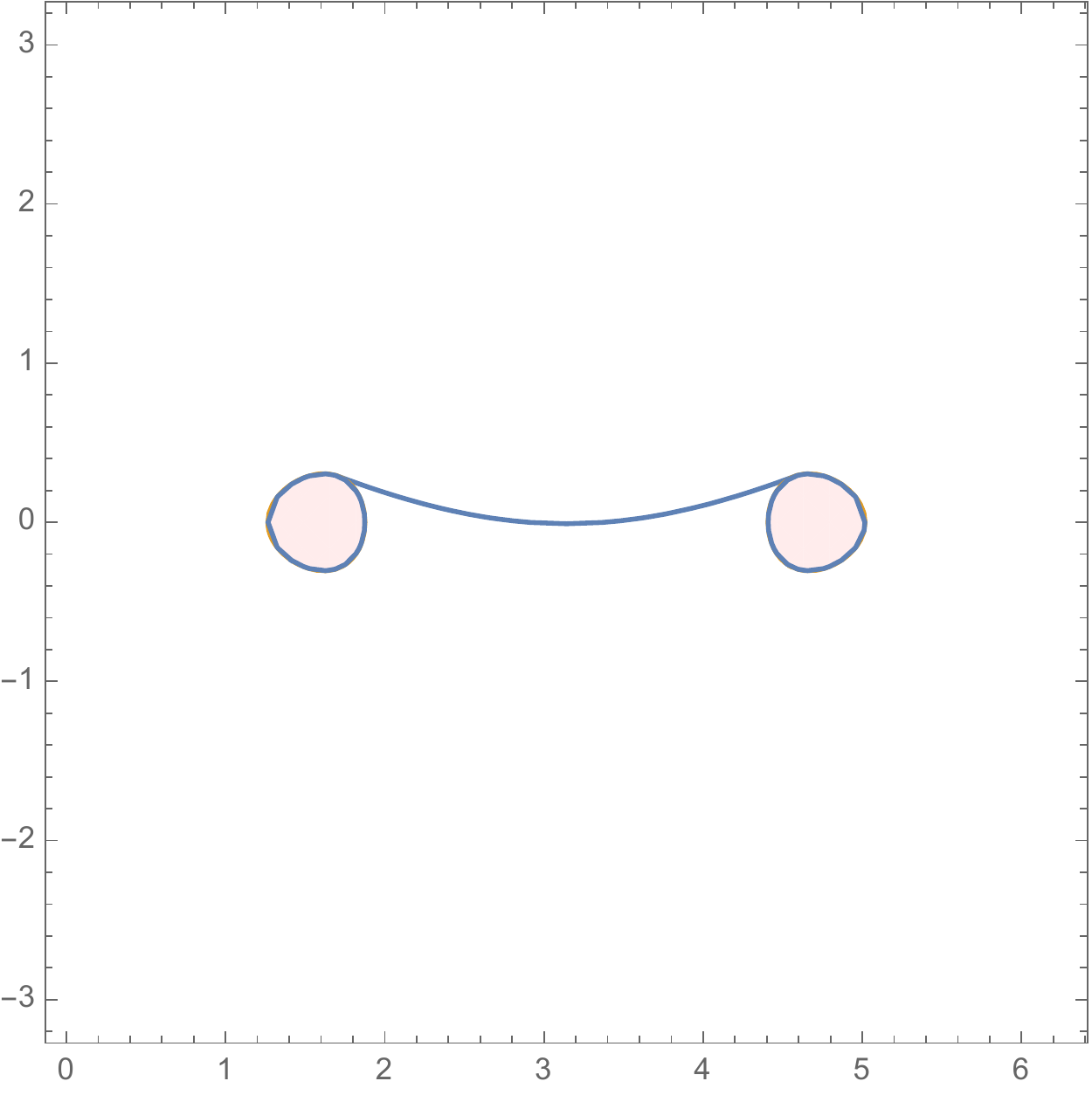}};
   \node (6) at (4,-8) {\includegraphics[width=10em]{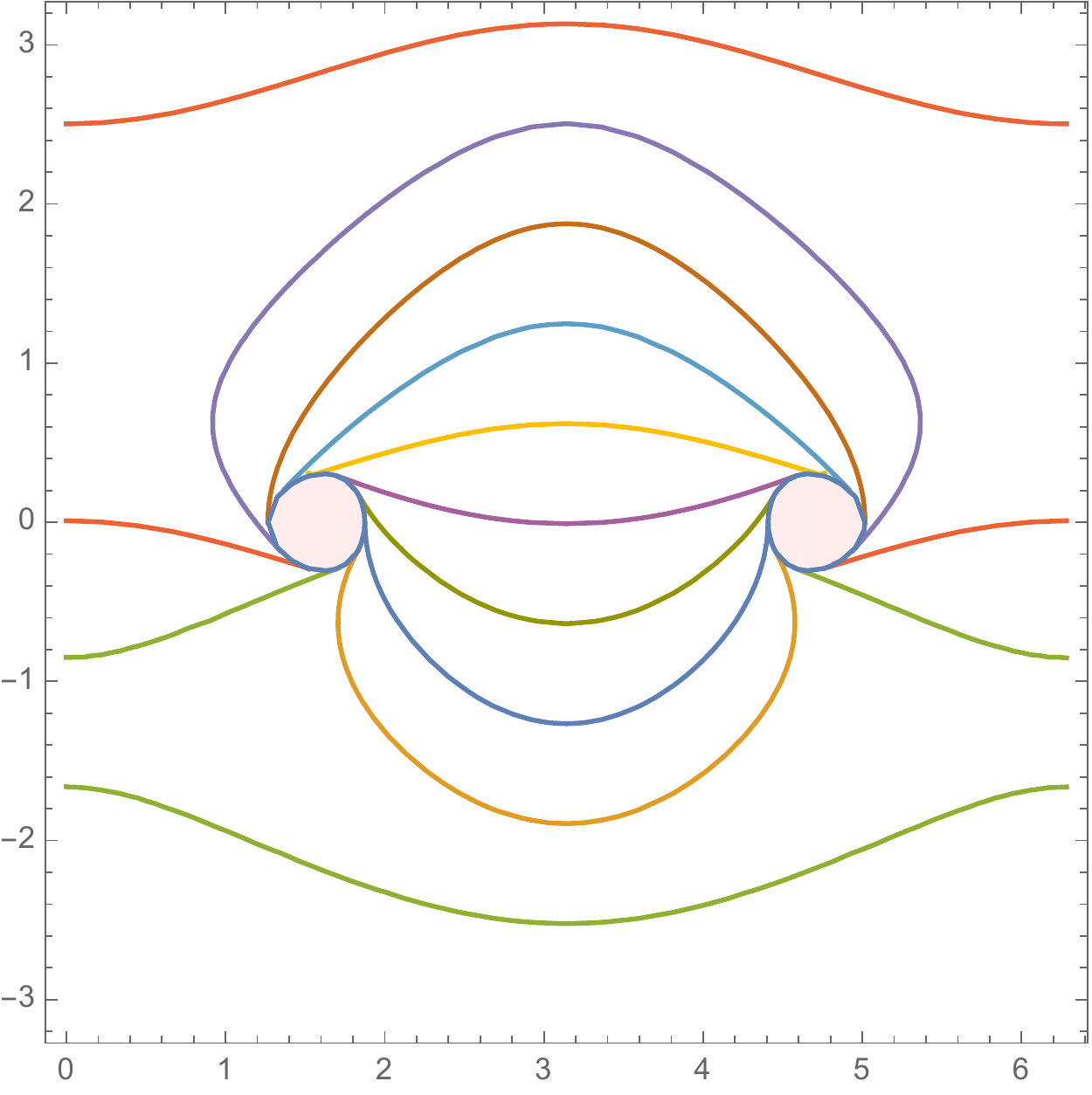}};

   \node at (.8,1.3) {$2\theta_+ = 0$};
   \node at (4.8,1.3) {$2\theta_+ = \frac{2\pi}{5}$};
   \node at (.8,1.3-4) {$2\theta_+ = \frac{4\pi}{5}$};
   \node at (4.8,1.3-4) {$2\theta_+ = \frac{6\pi}{5}$};
   \node at (.8,1.3-8) {$2\theta_+ = \frac{8\pi}{5}$};

   \begin{scope}[shift={(4,-8)}]

    \draw[->] (.78,-.22) arc (-90:250:.3);
    \draw[->] (-.68,-.22) arc (270:-50:.3);

    \node at (1.4,.5) {$\theta_+$};
    \node at (-1.2,.5) {$\theta_+$};
    
   \end{scope}
   
  \end{tikzpicture}
 \end{center}
 \caption{The parameter dependence of the Fermi arc which is the finite energy slice $\epsilon(p) = 0.3$ of the edge state energy spectrum with $2 \theta_+ = 0, 2\pi/5,4\pi/5,6\pi/5,8\pi/5$ for positive $\beta$. The horizontal and vertical axes are for $p_1$ and $p_2$. The shaded region shows the bulk spectrum. The last panel shows Fermi arcs with various parameter $\theta_+ \in [0,\pi)$.}
 \label{fig:Fermi_arc2}
\end{figure}
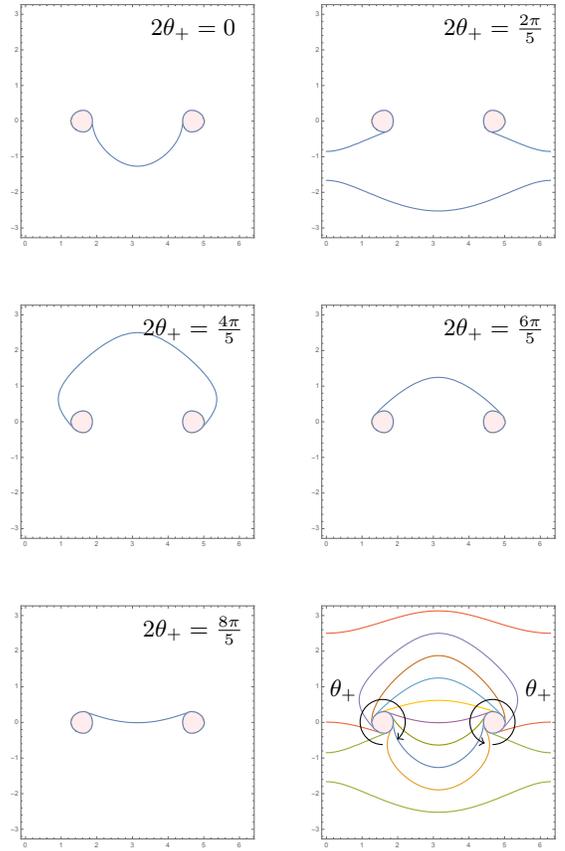

Assuming that the wavefunction is given by
\begin{align}
 \psi_{n_3} = \beta^{n_3-1} \psi_1
\end{align}
with the real parameter $|\beta| \le 1$, the eigenvalue equation \eqref{eq:eigen_prob_lat} is equivalent to
\begin{align}
 D \psi_{n_3} = 0
 \label{eq:zero_mode_eq}
\end{align}
where
\begin{align}
 D = 
 - \frac{i}{2}
 \begin{pmatrix}
  \beta^2 - 2 i \epsilon(p) \beta - 1  & 2 i \Delta(p)^\dag \beta \\
  - 2 i \Delta(p) \beta & \beta^2 + 2 i \epsilon(p) \beta - 1
 \end{pmatrix}
 \, .
\end{align}
To obtain a non-trivial solution to this zero mode equation, we asign the condition $\det D = 0$ which yields
\begin{align}
 \beta^2 & =
 1 + 2 (|\Delta(p)|^2 - \epsilon(p)^2)
 \nonumber \\
 & \quad
 -
 2 \sqrt{(|\Delta(p)|^2 - \epsilon(p)^2) (|\Delta(p)|^2 - \epsilon(p)^2 + 1)}
 \, .
\end{align}
There are two solutions for $\beta \ge 0$ and $\beta \le 0$.
We remark that these two possibilities correspond to the doublers at $p_3 = 0$ and $\pi$ in the momentum space.

Then, together with the boundary condition \eqref{eq:bc_lat3D2}, the zero mode equation \eqref{eq:zero_mode_eq} gives
\begin{align}
 D
 \begin{pmatrix}
  1 \\ -e^{2 i \theta_+}
 \end{pmatrix}
 = 0
 \, .
\end{align}
Since $\beta \in \mathbb{R}$, we obtain
\begin{align}
 \epsilon(p) & =
 - \cos 2\theta_+ \operatorname{Re}\Delta(p)
 - \sin 2\theta_+ \operatorname{Im}\Delta(p)
 \, , \\
 \tilde{\alpha}(p)
 %\frac{1}{2} \left( \beta^{-1} - \beta \right)
 & =
 \sin 2\theta_+ \operatorname{Re}\Delta(p)
 - \cos 2\theta_+ \operatorname{Im}\Delta(p)
 \, ,
\end{align}
which is rewritten as a matrix form
\begin{align}
 \begin{pmatrix}
  \epsilon(p) \\ \tilde{\alpha}(p)
 \end{pmatrix}
 = - 
 \begin{pmatrix}
  \cos 2 \theta_+ & \sin 2 \theta_+ \\
  - \sin 2 \theta_+ & \cos 2 \theta_+
 \end{pmatrix}
 \begin{pmatrix}
  \operatorname{Re}\Delta(p) \\
  \operatorname{Im}\Delta(p)
 \end{pmatrix}
 \, ,
\end{align}
where we define $\tilde{\alpha}(p) := \frac{1}{2} \left( \beta^{-1} - \beta \right)$, and from \eqref{eq:Delta}, the real and imaginary parts of $\Delta(p)$ are given by
\begin{align}
 \operatorname{Re}\Delta(p)
 & = \cos p_1 - \cos p_2 + c
 \, , \\
 \operatorname{Im}\Delta(p)
 & = \sin p_2
 \, .
\end{align}
Comparing with the continuum theory \eqref{rotea}, now the situation is parallel under the replacemtnt
\begin{align}
 (p_1, p_2, \alpha(p)) \ \longrightarrow \
 \left(
 \operatorname{Re}\Delta(p), \operatorname{Im}\Delta(p), \tilde{\alpha}(p)
 %\frac{1}{2}(\beta^{-1} - \beta)
 \right)
 \, .
\end{align}
Fig.~\ref{fig:edge_disp_lat3D} shows the boundary parameter dependence of the bulk and edge state dispersions.
The edge state spectrum has a support only where the normalizability conditioin is satisfied $|\beta| \le 1$. %$\alpha(p)$ is positive.
As mentioned before, there are two solutions corresponding to positive and negative $\beta$.
We focus on the positive solution in the following.
When we change the parameter $\theta_+$, the edge state spectrum rotates around the Weyl points.
The orientation, that is, how the edge state spectrum winds, depends on the chirality of the Weyl nodes.
This result is consistent with the continuum theory analysis in particular in the vicinity of the Weyl points.

To see the parameter dependence more explicitly, let us take the constant energy section of the spectrum, which yields the Fermi arc, shown in Figs.~\ref{fig:Fermi_arc1} and~\ref{fig:Fermi_arc2}.
This shows that the parameter characterizing the boundary condition $\theta_+$ plays a role of the rotation angle of the Fermi arc, as studied in continuum theory.
In the present case of the lattice models, 
the Fermi arc ends on the Weyl points and have a finite support in the momentum space.
Such a behavior of the Fermi arc has been experimentally observed, for example, in the transition metal pnictide family~\cite{Liu:NM2015}.

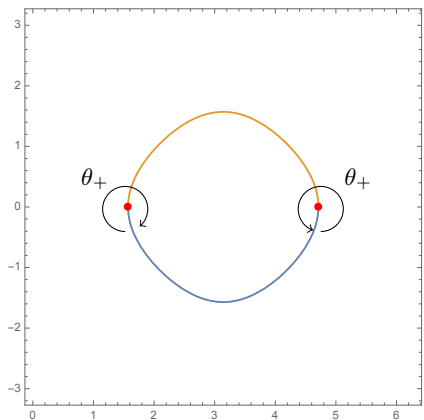
\begin{figure}[t]
 \begin{center}
  \begin{tikzpicture}

   \node at (0,0) {\includegraphics[width=17em]{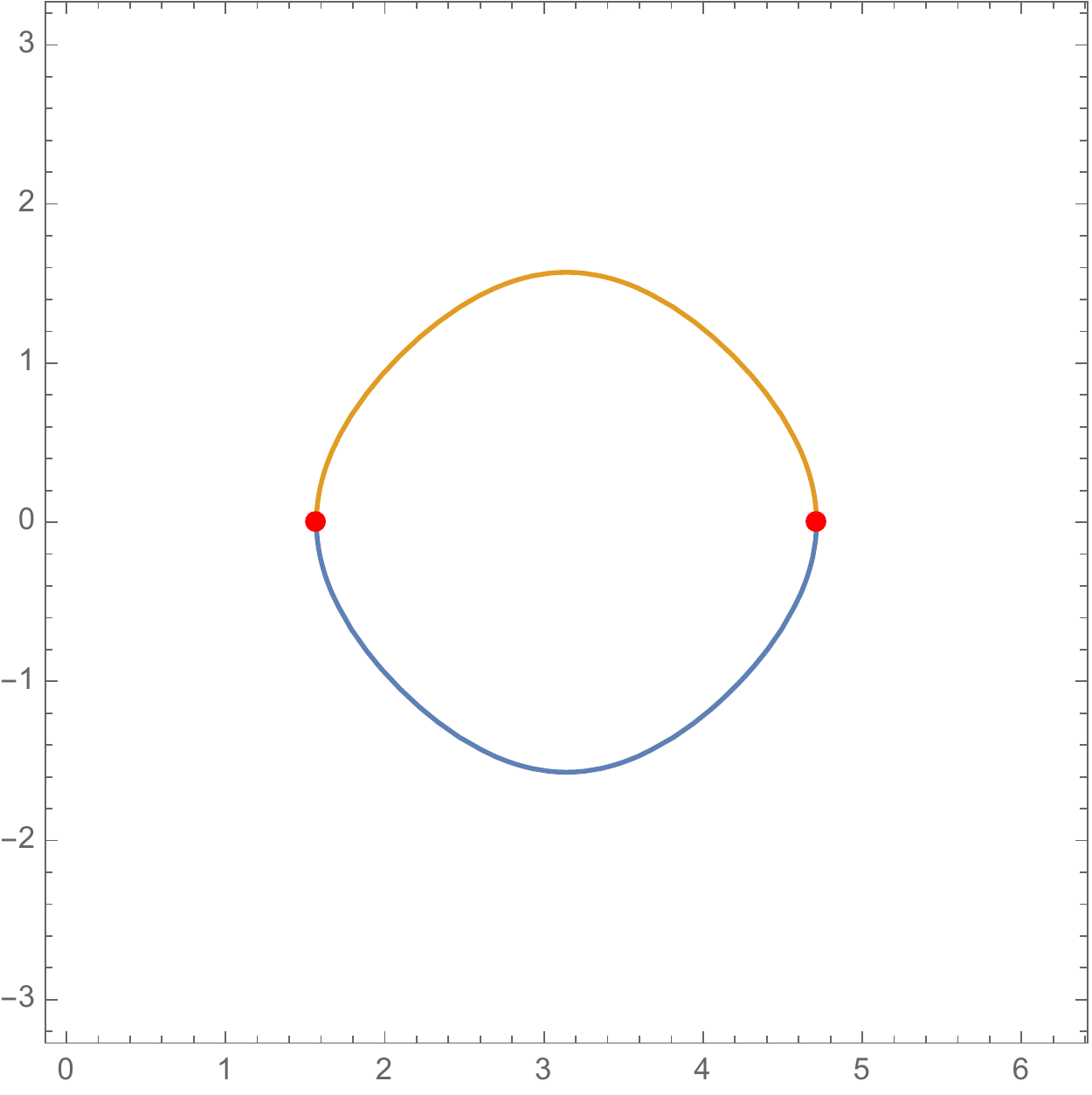}};

   \begin{scope}

    \draw[->] (1.4,-.2) arc (-90:250:.3);
    \draw[->] (-1.2,-.2) arc (270:-50:.3);

    \node at (1.9,.5) {$\theta_+$};
    \node at (-1.6,.5) {$\theta_+$};
    
   \end{scope}
   
  \end{tikzpicture}
 \end{center}
 \caption{The Fermi arcs for positive $\beta$ (blue) and negative $\beta$ (orange) solutions at $2\theta_+=0$. The rotation orientation depending of the boundary condition $\theta_+$ is the same for both cases.}
 \label{fig:arc_both}
\end{figure}

Let us comment on the Fermi arc behavior corresponding to the negative $\beta$ solution.
In this case the Fermi arc appears in the region $\tilde{\alpha}(p)<0$, which is complement to $\tilde{\alpha}(p)>0$ for the positive $\beta$ solution.
In addition, the rotation orientation for $\beta<0$, depending on the boundary condition $\theta_+$, is the same as that for $\beta>0$, as shown in Fig.~\ref{fig:arc_both}, and thus the winding number is also the same for both cases.
This implies that these two contributions from the positive and negative $\beta$ solutions are not canceled with each other.

\subsection{Reduction to 2D class A system}

As mentioned in Sec.~\ref{sec:2D_red} for continuum theory, the 3D Weyl semimetal system is translated to 2D class A system using the dimensional reduction.
Let us try this reduction to apply the systematic study on a lattice, and see how the topological edge state behaves under the generic boundary condition.

\begin{figure}[t]
 \begin{center}
  \includegraphics[width=12em]{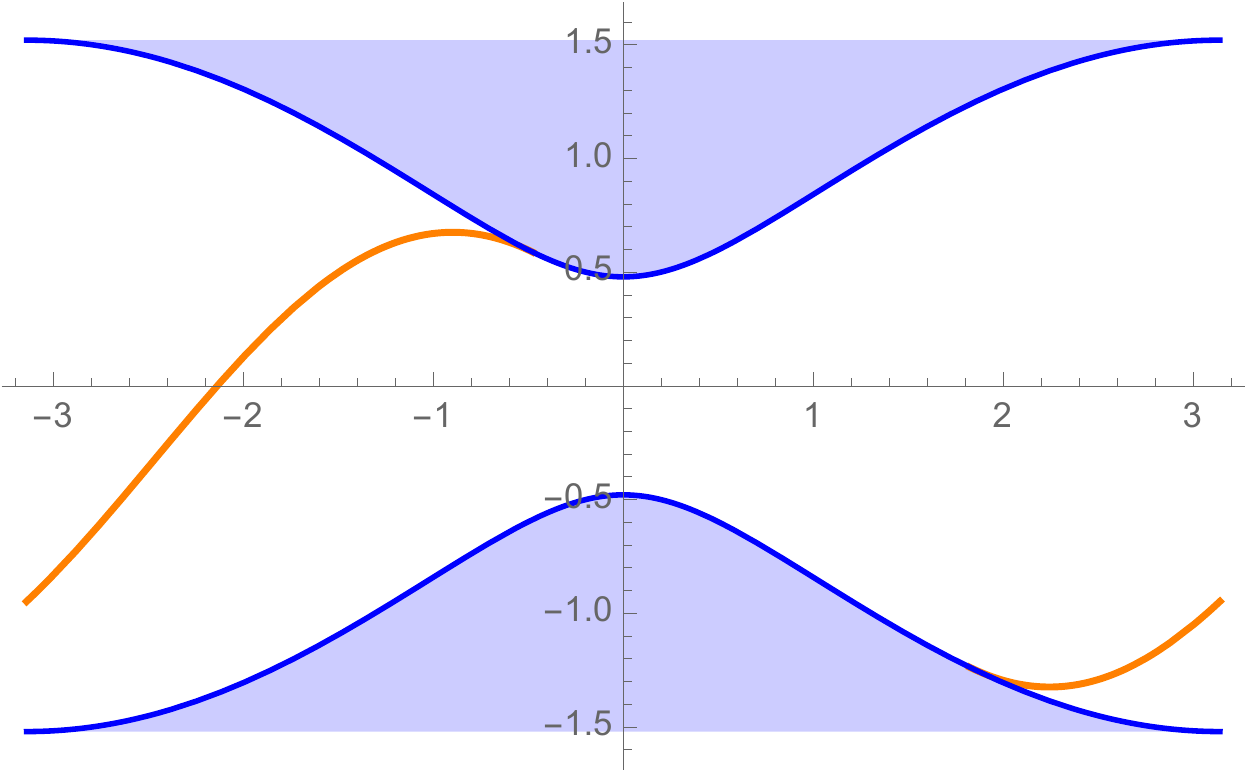} \quad
  \includegraphics[width=12em]{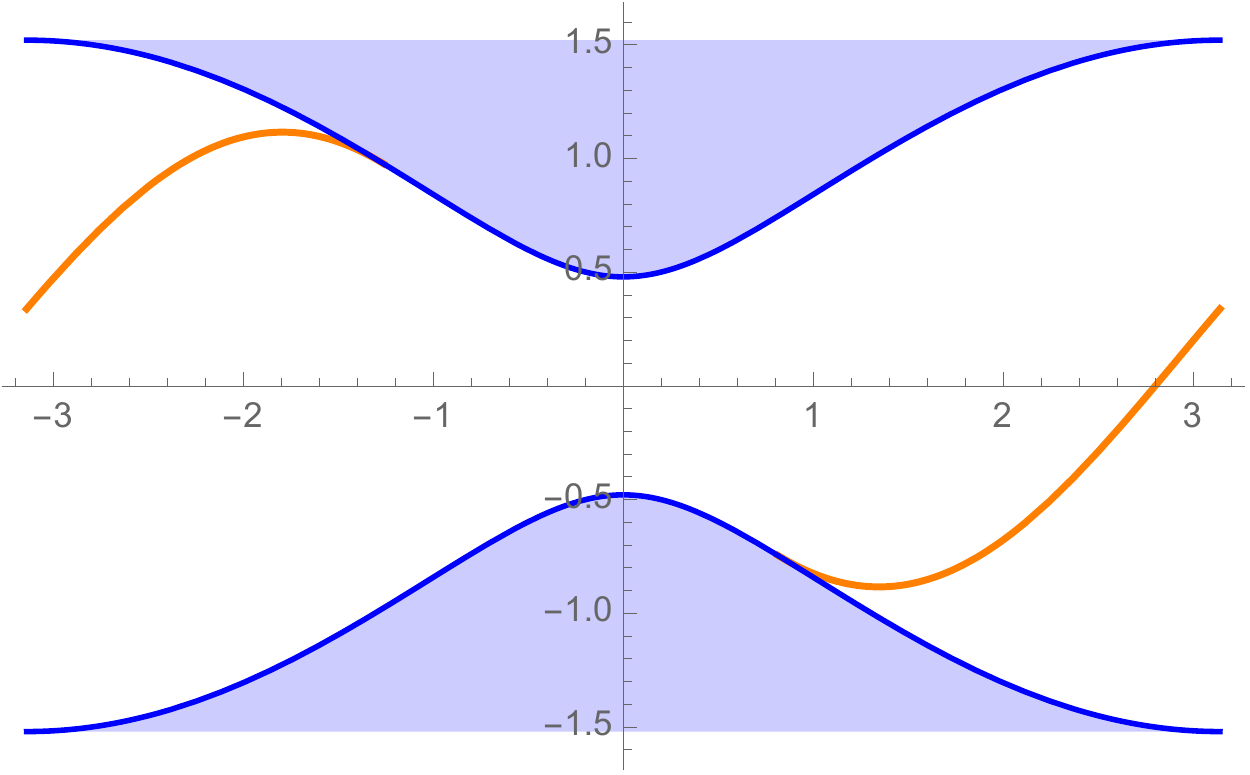} \\[1em]
  \includegraphics[width=12em]{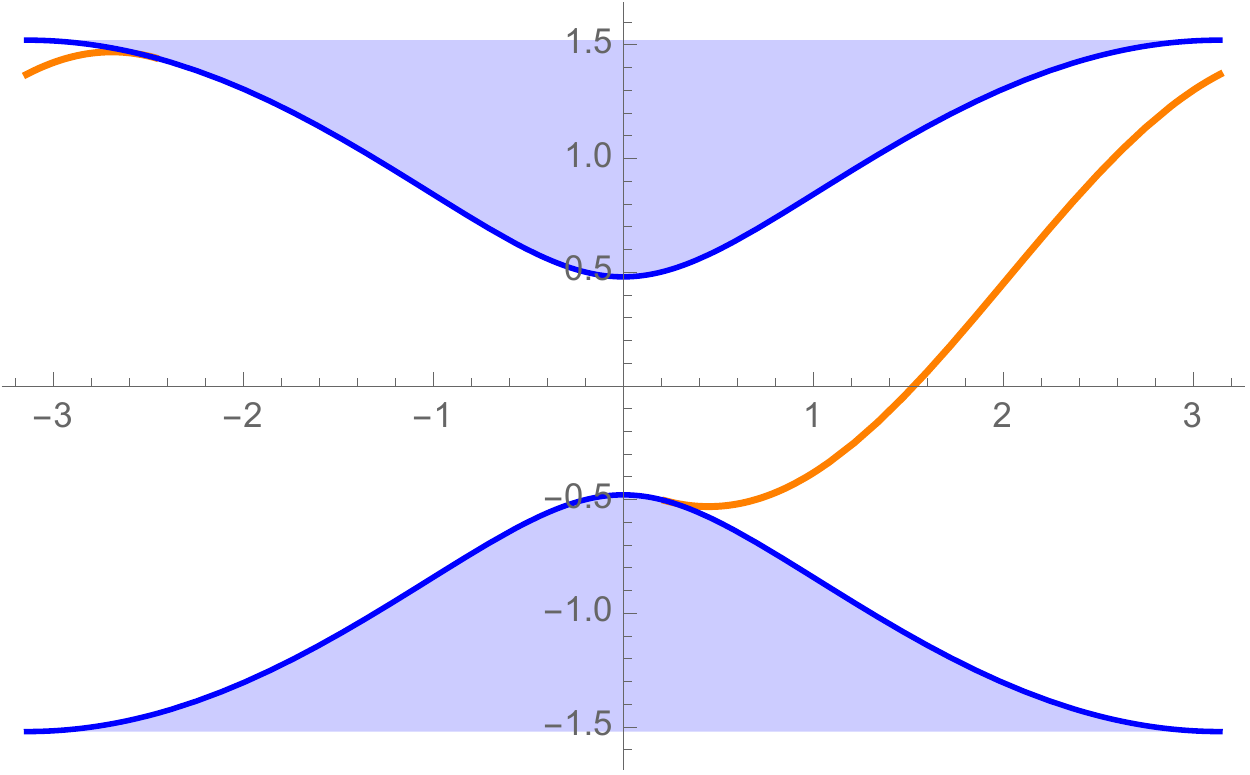} \quad
  \includegraphics[width=12em]{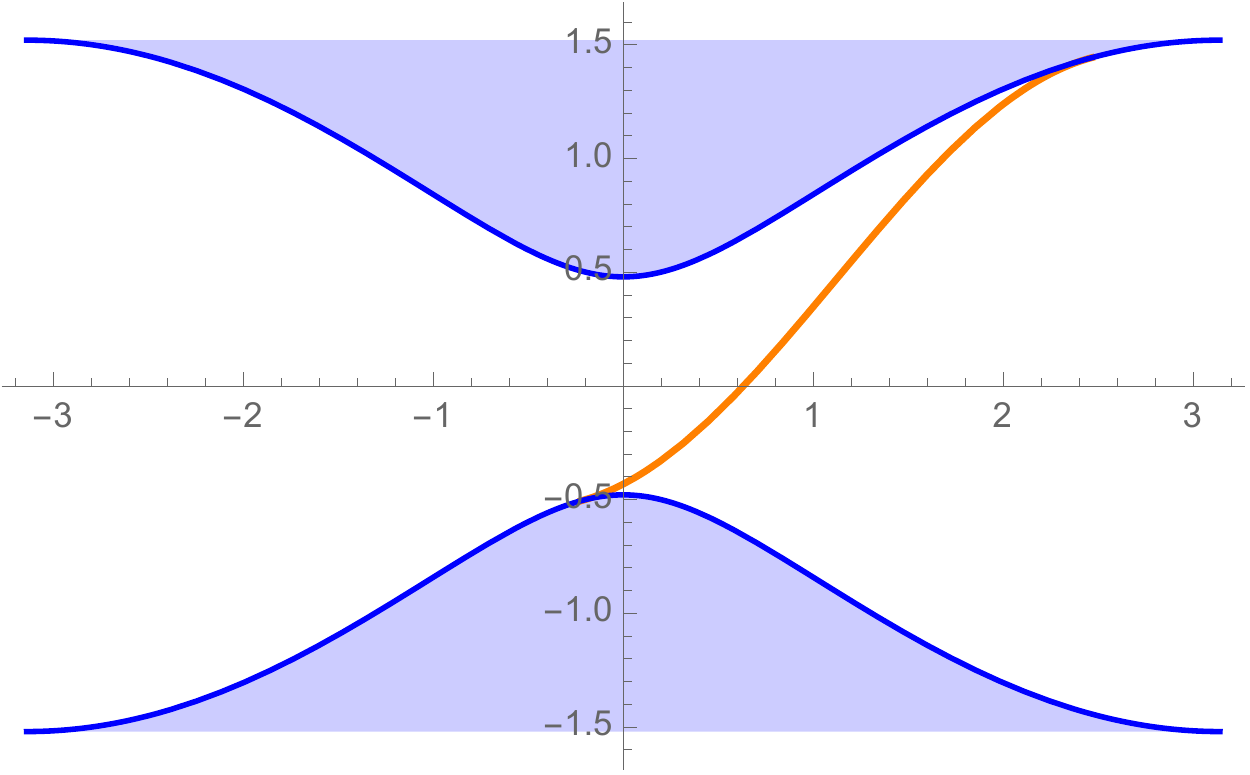} \\[1em]
  \includegraphics[width=12em]{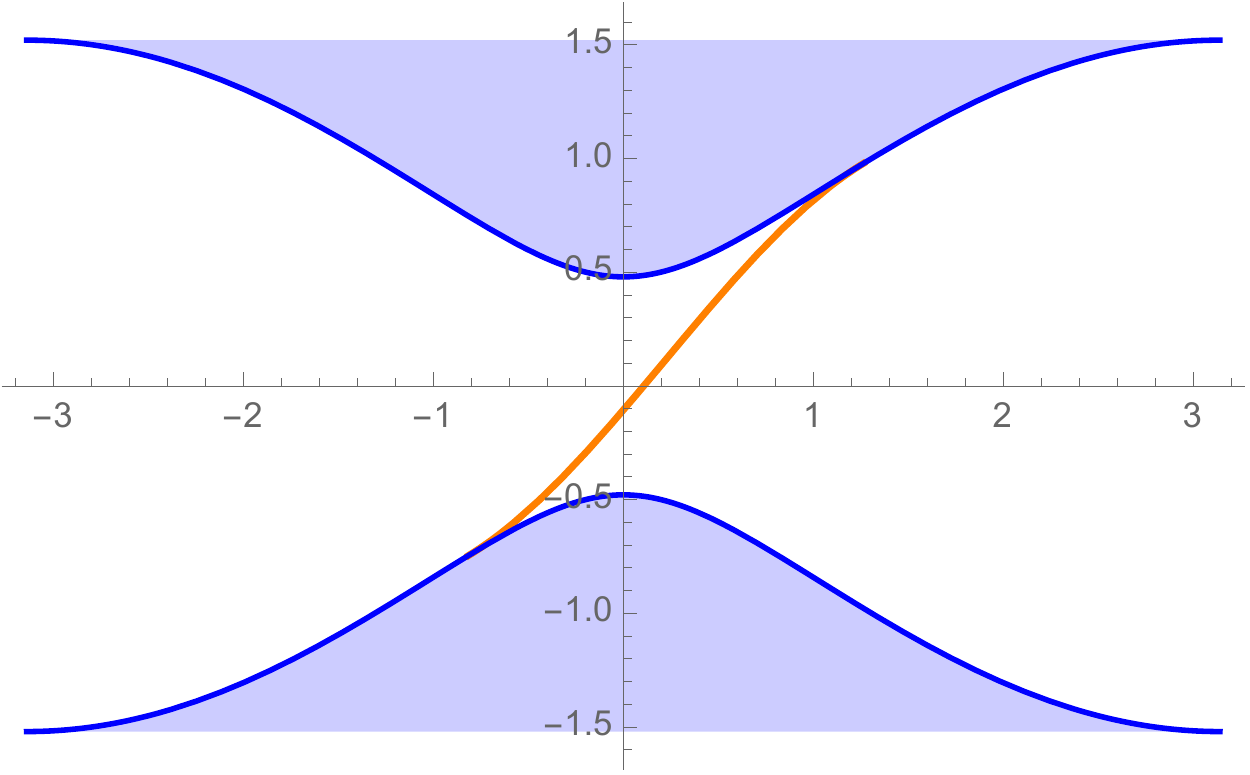} \quad
  \includegraphics[width=12em]{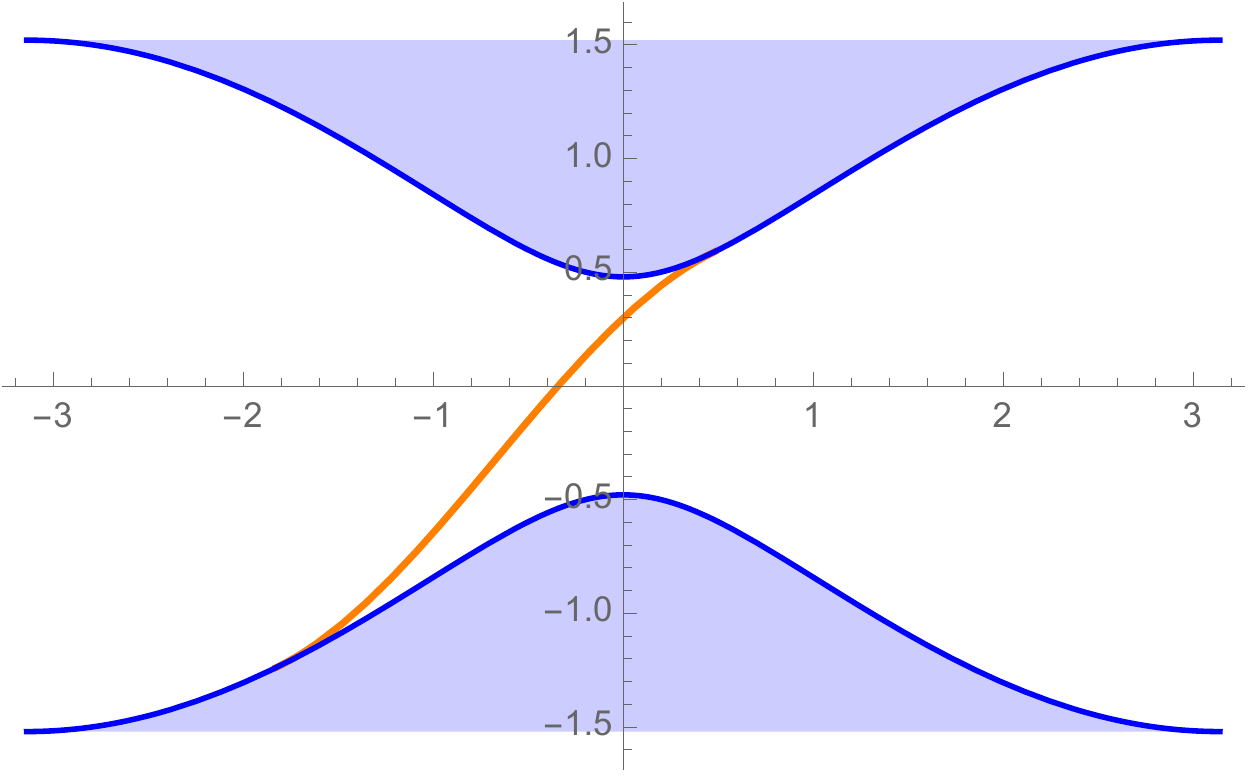}
 \end{center}
 \caption{The bulk and edge state dispersions from $p_1$-direction reduction with $c=1$, $m_1 = \pi/2 + 0.5$ and the boundary condition parameter $\theta_+= 2\pi/7,4\pi/7,6\pi/7,8\pi/7,10\pi/7,12\pi/7$ for positive $\beta$. The horizontal and vertical axes are for $p_2$ and $\epsilon$.}
 \label{fig:bulk_edge_disp_2Db}
\end{figure}

\begin{figure}[t]
 \begin{center}
  \includegraphics[width=12em]{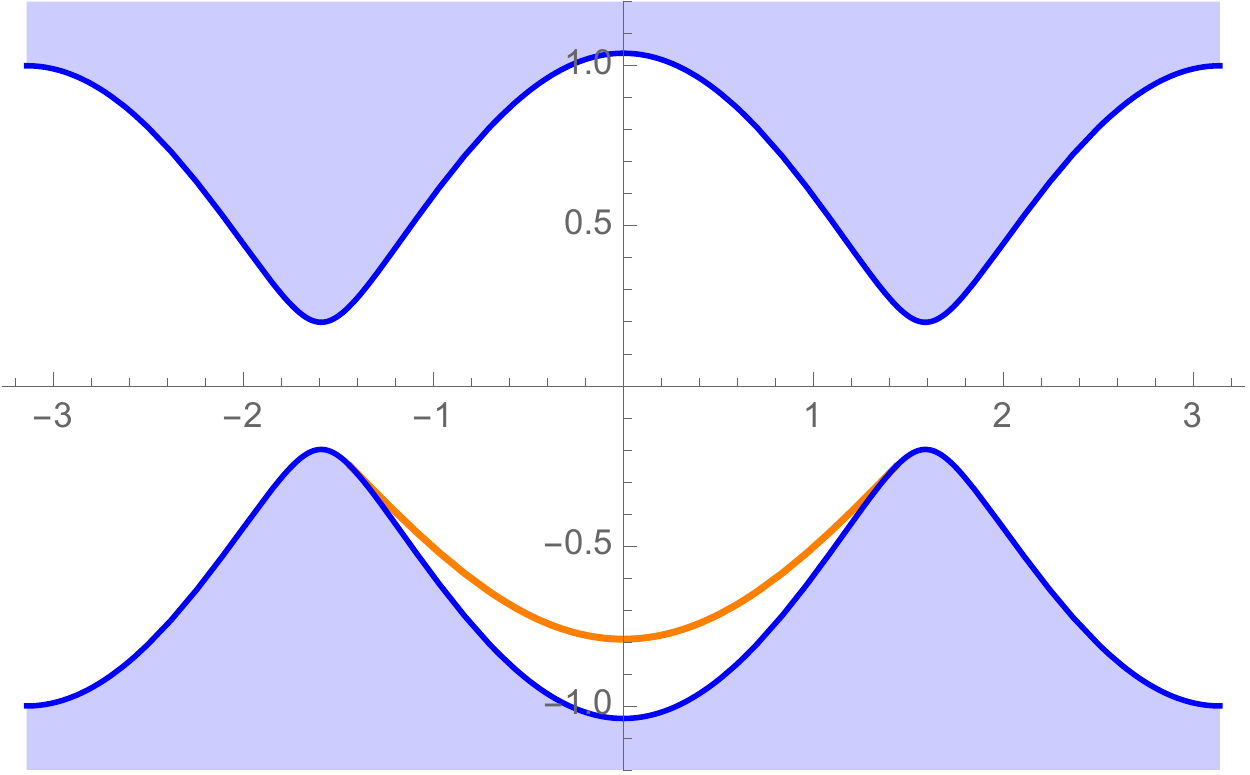} \quad
  \includegraphics[width=12em]{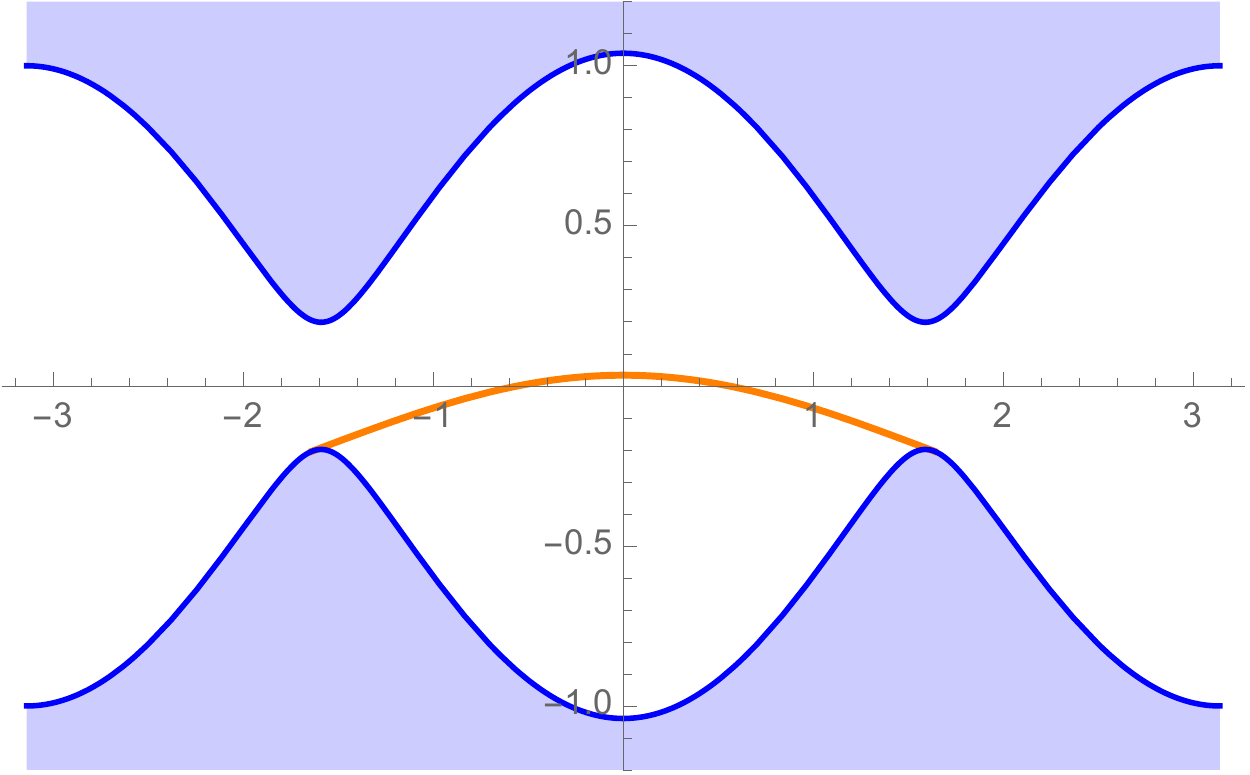} \\[1em]
  \includegraphics[width=12em]{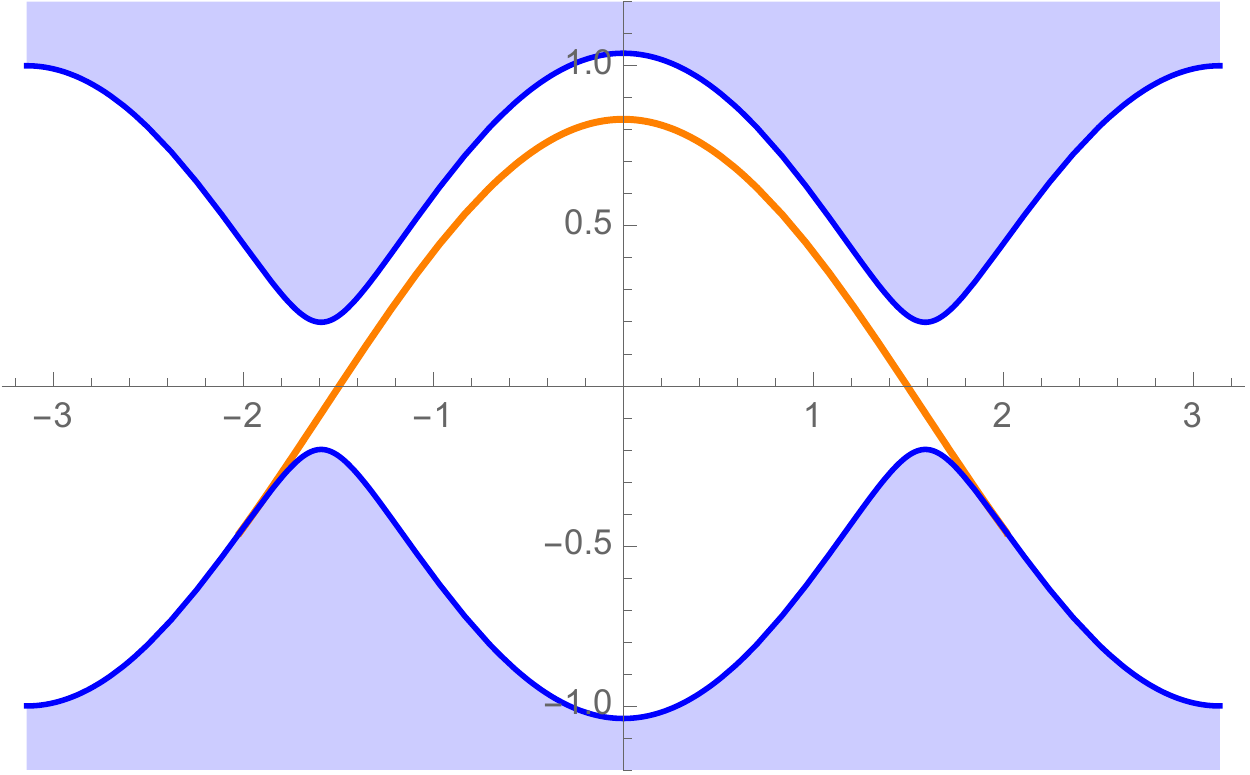} \quad
  \includegraphics[width=12em]{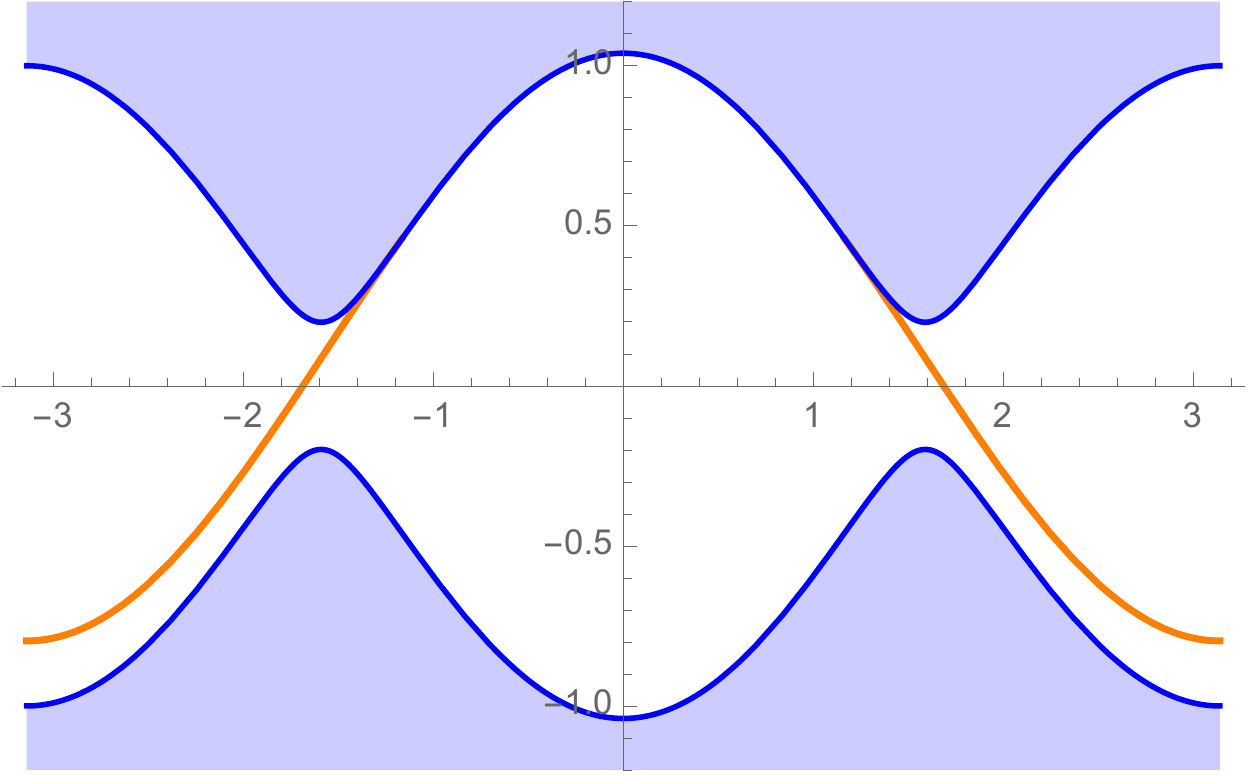} \\[1em]
  \includegraphics[width=12em]{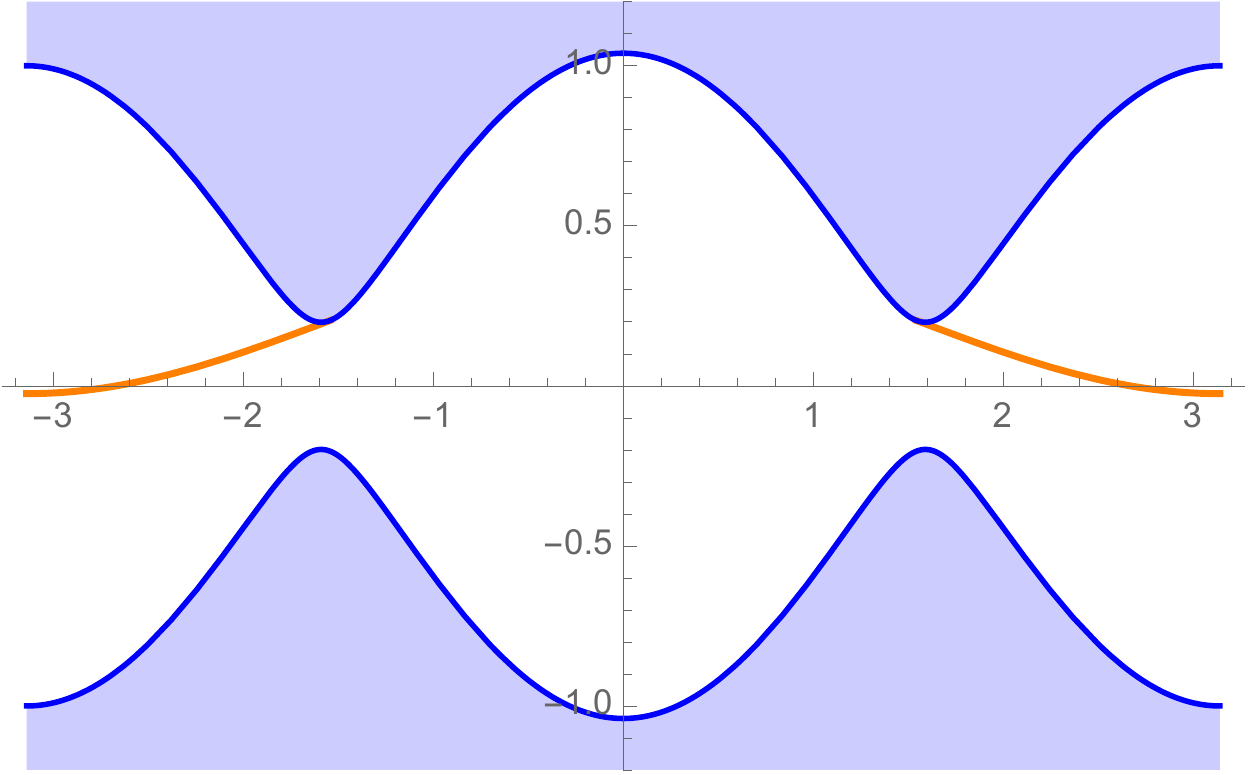} \quad
  \includegraphics[width=12em]{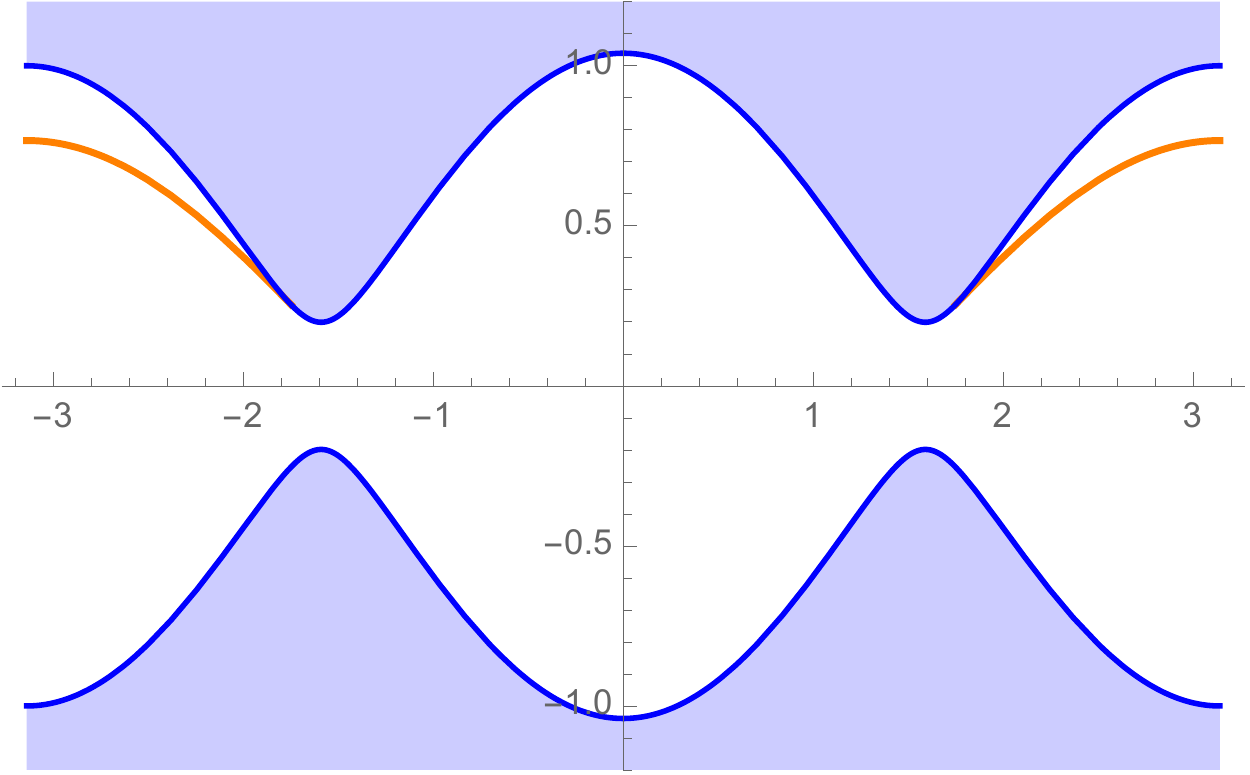}
 \end{center}
 \caption{The bulk and edge state dispersions from $p_2$-direction reduction with $c=1$, $m_2 = 0.2$ and the boundary condition parameter $\theta_+= 2\pi/7,4\pi/7,6\pi/7,8\pi/7,10\pi/7,12\pi/7$ for positive $\beta$. The horizontal and vertical axes are for $p_1$ and $\epsilon$.}
 \label{fig:bulk_edge_disp_2D}
\end{figure}

Replacing the momentum $p_1$ or $p_2$ with the constant mass parameter $m$ in the 3D Hamiltonian \eqref{eq:3D_Bloch_Ham}, we obtain the 2D class A system
\begin{align}
 \mathcal{H}_1(p) & =
 \sigma_1 (\cos m_1 - \cos p_2 + c ) 
 + \sigma_2 \sin p_2 + \sigma_3 \sin p_3
 \, , \\
 \mathcal{H}_2(p) & =
 \sigma_1 (\cos p_1 - \cos m_2 + c ) 
 + \sigma_2 \sin m_2 + \sigma_3 \sin p_3
 \, .
 \label{eq:3D_Bloch_Ham}
\end{align}
This is the dimensional reduction along the $p_{1,2}$-direction.
Putting the boundary at $n_3 = 1$, we can apply the same self-conjugacy argument to obtain the boundary condition as the 3D system~\eqref{eq:bc_lat3D}.
Thus the edge state spectrum is given by
\begin{align}
 \begin{pmatrix}
  \epsilon \\ \tilde{\alpha}
 \end{pmatrix}
 & = -
 \begin{pmatrix}
  \cos 2 \theta_+ & \sin 2 \theta_+ \\
  - \sin 2 \theta_+ & \cos 2 \theta_+
 \end{pmatrix}
 \begin{pmatrix}
  \operatorname{Re}\Delta(p_{1,2}=m_{1,2}) \\
  \operatorname{Im}\Delta(p_{1,2}=m_{1,2})
 \end{pmatrix}
% \begin{pmatrix}
%  \cos p_1 - \cos m + c \\ \sin m
% \end{pmatrix}
 \, .
\end{align}
Figs.~\ref{fig:bulk_edge_disp_2Db} and \ref{fig:bulk_edge_disp_2D} show the bulk and edge state dispersions depending on the boundary condition parameter $\theta_+$, which is again consistent with the continuum theory.
Such a dependence of the boundary condition was recently predicted to be observed in monolayer silicene/germanene/stanene nanoribbons~\cite{hattori2016edge}.

%%%%%%%%%%%%%%%%%%%%%%%%%%%%%%%%%%%%%

\section{The bulk-edge correspondence}\label{sec:bec}

%%%%%%%%%%%%%%%%%%%%%%%%%%%%%%%%%%%%%
In this section, we 
study the relation between the bulk and the edge states, and 
check the bulk-edge correspondence. 
We will see that for the most generic boundary conditions, the bulk-edge correspondence 
for the 2D topological phase works perfectly. 
The bulk-edge correspondence \cite{Jackiw:1975fn,hatsugai1993chern,Wen:2004ym}
for topological insulators is well-known,
while that for 3D Weyl semimetals has been understood in a way through a dimensional reduction to 2D.
We claim here that the bulk-edge correspondence for the 3D Weyl semimetals
can be defined as follows: 
the topological number of the bulk counts the chirality of the Weyl fermions, while
the topological number of the edge 
is defined through the orientation of the Fermi arcs attached to the Weyl cone. 
This is based on our analysis on most general boundary conditions of the Weyl semimetals.

In the following, first we investigate the case of the two dimensions, then later we discuss the case of the three 
dimensions, the Weyl semimetals.

%On three dimensions, which describes a Weyl semimetal, it is somehow ambiguous but 

\subsection{2D topological phases with the most general boundary conditions}

For the bulk-edge correspondence for the 2D topological insulator of class A, 
the formula is given by 
\begin{align}\label{BEC}
	k=n_+-n_-,
\end{align}
where $k$ is the bulk topological number (the TKNN number), and $n_{\pm}$ counts the number of
right/left moving edge states. 
%We want to see the meaning of this equation in two and three dimensions and how they are related by a dimensional reduction. %How?
%In two dimensions, which describes a class AIII topological insulator, the relation between bulk and edge is very clear: the bulk topological number corresponds to TKNN number and the one defined by edge states is just the numbers of right/left moving edge states. 

\begin{figure}[h!]
	\begin{center}
	 \includegraphics[scale=0.25]{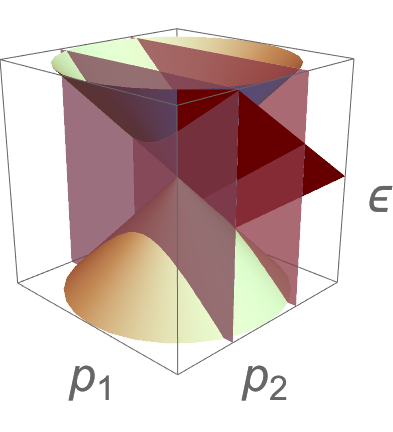}
	 \includegraphics[scale=0.25]{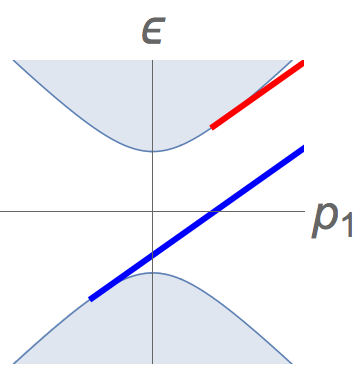}
	 \includegraphics[scale=0.25]{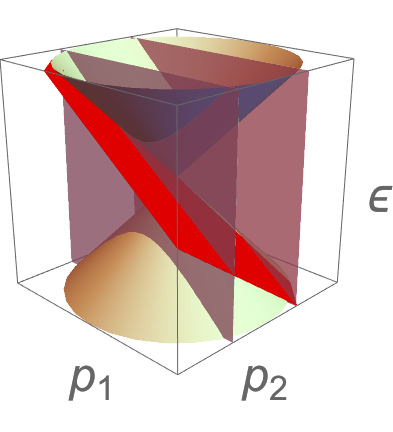}
	 \includegraphics[scale=0.25]{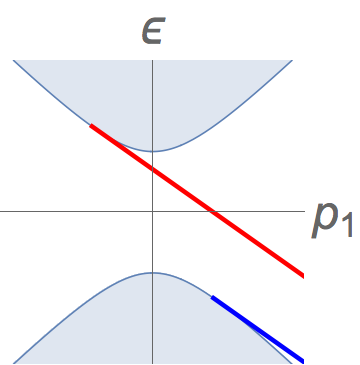}	 
	 \caption{3D and 2D edge states with respectively $2\theta_+=\frac{3}{4}\pi,~ \pi/4$. On the right hand side, blue line is the edge state with $m>0$ and red line is the edge state with $m<0$.}
	 \label{fig:3D-2D}
	\end{center}
\end{figure}

In two dimensions, both sides of \eqref{BEC} should be understood as the difference under sign flip of the 
parameter $m$ for each cone.
The topological number $k$ is defined as the difference of the TKNN number when $m$ changes its sign:
\begin{align}
	k=\nu(m>0)-\nu(m<0).
\end{align} As for the topological number of the gapless edge states, in our single fermion problem, we choose it as the sign of
\begin{align}
	\frac{\partial \epsilon}{\partial p_1}.
\end{align}
When it has a plus sign, we denote is as $n_+=1$, and when it has a minus sign, we denote it as $n_-=1$. 

Using these definitions, we check various cases with different values of $\theta_+$, and 
we find that all are consistent with the bulk-edge correspondence (\ref{BEC}).
Here, as an illustration, we show only two typical examples.
In Fig~\ref{fig:3D-2D}, when $2\theta_+=\frac{3}{4}\pi$, we have
\begin{align*}
	n_+(m>0)=1, ~n_-(m>0)=0;
	\\
	n_+(m<0)=0, ~n_-(m>0)=0,
\end{align*}
and 
when $2\theta_+=\pi/4$, we have
\begin{align*}
	n_+(m>0)=0, ~n_-(m>0)=0;
	\\
	n_+(m<0)=0, ~n_-(m>0)=1.
\end{align*}
The both cases have 
\begin{align}
	 \Delta  n_+ - \Delta n_-=1.
\end{align}
On the other hand, the TKNN number is calculated as
\begin{align*}
	\nu(m>0)=1/2,	
	\hspace{3mm}
%\end{align*}
%\begin{align*}
	\nu(m<0)=-1/2,	
\end{align*}
so we find consistency in these two examples:
\begin{align}
	k= \Delta  n_+ - \Delta n_-.
\end{align}
In this way, for all possible values of $\theta_+$, we show the bulk-edge correspondence.
This means that the correspondence is true for any consistent boundary conditions in 2 dimensions.

\begin{figure}[t]
	\begin{center}
	 \includegraphics[scale=0.25]{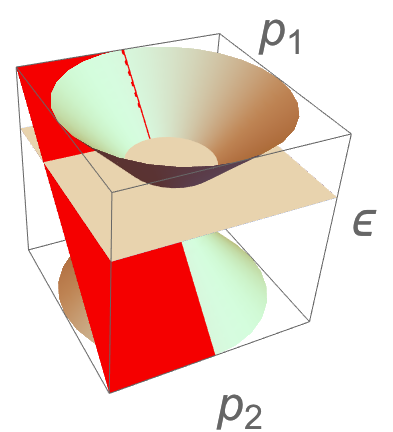}
	 \includegraphics[scale=0.25]{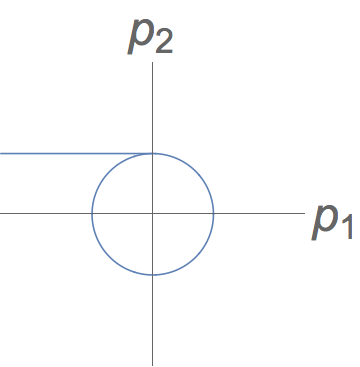}
	 \includegraphics[scale=0.25]{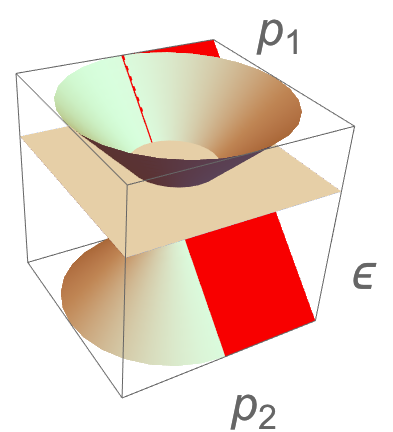}
	 \includegraphics[scale=0.25]{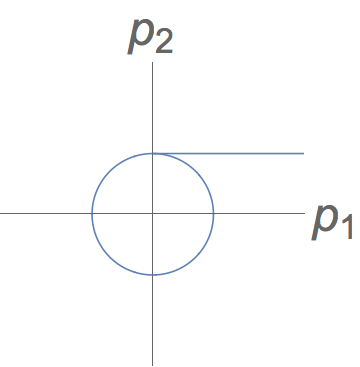}	 
	 \caption{
	 How to count the number of edge states from the orientation of the Fermi arcs. 
	 Top: a Fermi arc emanates from the Weyl node with a positive chirality $K=1$, which is a counter-clockwise.
	 Bottom: the case of the opposite chirality, $K=-1$.
	 }
	 \label{fig:FA}
       \end{center}
\end{figure}

%%%%
\subsection{3D Weyl semimetals with the most general boundary conditions}

Let us turn to the case of the 3 dimensions, the Weyl semimetals.
In three dimensions, the topological number $K$ is defined by the wrapping number of a map
$b_i(p_j)$ which shows up in the Hamiltonian
%\begin{align}
%	\{ b_1,b_2,b_3\}  \to \{ p_1, p_2, p_3\}
%\end{align}
%where 
\begin{align}
	\mathcal{H}=b_i(p_j)\sigma_i.
\end{align}
Our Hamiltonian (\ref{Ham}) is given by  $b_i=p_i$ and the Weyl node is at $p_i=0$.
Considering a two-sphere surrounding the Weyl node, we obtain 
\begin{align}
	K=1\in\pi_2(S^2).
\end{align}

We claim the bulk-edge correspondence for the 3D Weyl semimetal is given by
\begin{align}
K = N-\tilde{N}
\label{3DBE}
\end{align}
where $K$ is the topological number defined above.
We define $N$ and $\tilde{N}$ to count the numbers of edge states with independent orientations
with respect to the orientation of the bulk dispersion cone, as we will see below.
% once the orientation of the boundary surface is fixed. 

%To define the right/left moving modes of the edge states in the is not . 

To discuss the orientation, we have to view the bulk dispersion in the subspace $(p_1,p_2)$ since the
edge dispersion lives in that space.
%To relate the bulk topological number with edge states, we first notice that 
First, in  the $(p_1, p_2, \epsilon)$ space, we notice that 
all constant $p_3$ slices of the bulk states have the same orientation.
%: the projection of $\hat{p}_1\times\hat{p}_2$ on $\epsilon$ direction always has the same sign once the Hamiltonian is determined and this is because of bulk topological number. 
Let us make further a slice at a constant positive energy $\epsilon$. The cross-section of the bulk dispersion
is a circle (see Fig.~\ref{fig:FA}). The orientation of the circle is definite due to the topological number (assuming $b_3=p_3$).

The constant energy slice of the edge dispersions defines the Fermi arcs.
Since generically the edge state dispersions are planes tangent to the bulk dispersions, 
the Fermi arcs share the same property.
The number $N$ counts the number of Fermi arcs which are tangential to the bulk dispersion circle and emanates
in a counter-clockwise orientation. On the other hand, the number $\tilde{N}$ counts that in a clockwise orientation
\footnote{
Note that 
$N$ and $\tilde{N}$ defined here are meaningful only with their associated Weyl cone.
For example, for an edge state connecting two Weyl cones, the numbers $N$ and $\tilde{N}$ cannot be defined in a uniformed way. 
}.
Our claim for the bulk-edge correspondence is that this orientation of the bulk circle remains the same for the edge
(the Fermi arcs).

Let us check this explicitly for two typical examples. In Fig.~\ref{fig:FA} we show 
the Hamiltonian (\ref{Ham}) with $\theta_+=0$, and the case of the Hamiltonian 
${\cal H}=-p_1\sigma_1 + p_2 \sigma_2 + p_3 \sigma_3$ with $\theta_+=0$.
%corresponding to a tangent vector field of the cone. 
The former case has $K=1$ as explained before, while the latter case has $K=-1$.
As we can see in Fig.~\ref{fig:FA}, it is obvious that 
we have $(N, \tilde{N})=(1,0)$ for the former case,
and $(N, \tilde{N})=(0,1)$ for the latter case.
So, they are consistent with our claim of the bulk-edge correspondence (\ref{3DBE}).

All the edge states in Fig.~\ref{Dispersion1} have the same $N$ and $\tilde{N}$ according to our definition: $(N, \tilde{N})=(1,0)$. 
So they are consistent again with (\ref{3DBE}). 
The examples in the lattice models we considered are shown to be consistent with the bulk-edge correspondence.
Note that Fermi arcs join Weyl nodes, and our counting works for each Weyl node. To be more precise,
each Fermi arc has two end points, and one end has $(N, \tilde{N})=(1,0)$ while the other end has 
$(N, \tilde{N})=(0,1)$. So the numbers are assigned to each end point of the Fermi arc.

%%%%%%%%%%%%%%%%%%%%%%%%%%%%%%%%%%%%%

\section{New topological structure and Berry phase}\label{sec:new_top_num}

%%%%%%%%%%%%%%%%%%%%%%%%%%%%%%%%%%%%%

In Sec.~\ref{sec:bc_3DWeyl}, we saw that the boundary condition of the 3D Weyl semimetals, as well as that of the 2D system, has only a single real parameter $\theta_+$. Since this $\theta_+$ is a new parameter
describing the system with a boundary, we can think of it as a coordinate in the theory space. 
Normally, for a given Hamiltonian, the theory space is spanned by parameters of the Hamiltonian.
From a topological viewpoint, the parameters are conserved momenta. The 3D Weyl semimetals
are of that category, and the parameter dependence of the Hamiltonian defines the 
bulk topological charge. Now, once we introduce the boundary, one of the momenta becomes ill-defined
and drops off from the list of the parameters. However, there shows up a set of new parameters
describing the boundary condition. From the analyses of this paper, we know that the new parameter
is only $\theta_+$. So, the most general parameter space of the 3D Weyl semimetals with a boundary in
the continuum limit is described by $(p_1,p_2,\theta_+)$.

To look for a novel topological structure of the system with a boundary, we study 
the wave function of the edge states, which depends only on the three parameters $(p_1, p_2,\theta_+)$.
The nontrivial topological structure can often be detected by a Berry phase in the parameter space. 
We will find that, for the present case, the only non-vanishing Berry connection is that for $\theta_+$,
and it provides us with a nontrivial Berry phase along a path in the parameter space.
%only $A_{\theta_+}$ is nonvanishing but it gives nonvanishing Berry phase along some path.

Before getting to the details, we first note that the important part is just the phase of the wave function,
to obtain a nonzero Berry connection.
Suppose that the phase of the wave function  does not depend on a parameter $\beta$. 
Then we find easily that
\begin{align}
	\psi^\dagger \frac{d}{d \beta}\psi
	=
	|\psi|^{\text{T}}\frac{d}{d \beta} |\psi|
	=
	(\frac{d}{d \beta}|\psi|^{\text{T}} )|\psi|.
\end{align}
Under this equality, the Berry connection associated with the parameter $\beta$ is
\begin{align*}
	A_{\beta}
	&=i\psi^\dagger \frac{d}{d \beta}\psi
	\\
	&=i|\psi|^{\text{T}}\frac{d}{d \beta} |\psi|
	\\
	&=i(\frac{d}{d \beta}(|\psi|^{\text{T}}|\psi|)-(\frac{d}{d \beta}|\psi|^{\text{T}} )|\psi|)
	\\
	&=0-A_{\beta},
\end{align*}
which means the vanishing of the Berry connection, $A_{\beta}=0$. 

Now, if we look at our general edge wave function \eqref{edgegeneral}, the phase does not depend on $p_1$ and $p_2$. Therefore, we conclude 
\begin{align}
A_{p_1}=A_{p_2}=0
\end{align}
in our generic parameter space.

However the phase of the wave function \eqref{edgegeneral} depends on $\theta_+$, and 
we can calculate the Berry connection as
\begin{align}
	A_{\theta_+}&=i\int dx^3~ \psi^{\dagger} \frac{d}{d\theta_+}\psi \nonumber
	\\
	&=i\int dx^3~\alpha e^{-2\alpha x^3}e^{2i\theta_+}\frac{d}{d\theta_+}e^{-2i\theta_+} \nonumber
	\\
	&=1.
	\label{At+}
\end{align}
Note here that the Berry connection for the edge state, which has $x^3$ dependence, needs in
its definition the integral over $x^3$ space so that the connection becomes Hermitian.
So, we have a nontrivial Berry connection along the parameter $\theta_+$, which dictates the
most general boundary condition of the 3D Weyl semimetals.

With the non-vanishing Berry connection at hand, let us study a possible topological charge.
The range of the parameter $\theta_+$ is, as was analyzed earlier, the period $0<\theta_+ \leq \pi$.
The point $\theta_+=0$ is identical with the point $\theta_+=\pi$, so it describes a circle
\footnote{
Note that we have chosen the phase of the 
wave function of the edge states \eqref{edgegeneral}
such that $\psi(\theta_+=0)=\psi(\theta_+=\pi)$.
}.
We consider a path going around this circle once, with some dependence on $p_1$ and $p_2$.
Let us calculate the Berry phase $\phi_{\rm B}$ 
along this path in the parameter space $p_1, p_2$ and $\theta_+$. 
Using \eqref{At+}, we obtain
\begin{align}
	\phi_{\rm B}
	&=\int_{\theta_+:~0 \to \pi} 
	\left[A_{p_1}dp_1+A_{p_2}dp_2+A_{\theta_+}d\theta_+\right]
	 \nonumber
	\\
	&=\pi.
\end{align}
This means that the edge state has a new topological charge, and its value is $1/2$.

In this calculation we considered a closed path in the $(p_1,p_2,\theta_+)$ space. 
Let us check whether the path exist or not. 
For the edge state to exist, we need to satisfy the normalizability condition
$\alpha>0$. This amounts to a nontrivial relation
\begin{align}
	\alpha=p_1\sin2\theta_+ -p_2\cos2\theta_+>0.
\end{align}
In changing $\theta_+$ from $0$ to $\pi$, it is necessary to choose a path in the $(p_1,p_2)$ space
so that the above inequality is satisfied. An example of such a path is given by
$(p_1,p_2) = c \, (\sin 2\theta_+, -\cos2\theta_+)$ with a positive constant $c$.

One may expect that the 2D case should have a similar topological number. Unfortunately, this is not the case.
Since $\alpha$ has to be positive, a constant $m$ gives a constraint on $\theta_+$:
\begin{align}
	\alpha=p_1\sin2\theta_+-m\cos2\theta_+>0.
	\label{alphacond2}
\end{align}
This has no solution for $p_1$ for a given $m$ and all possible $\theta_+$.
For example, for any positive $m$,  at $\theta_+=0$, there is no $p_1$ satisfying the inequality.
In the same manner, for any negative $m$, at $\theta_+=\pi/2$, there is no $p_1$.
So, the constancy of $m$ does not allow any path going from $\theta_+=0$ to
$\theta_+=\pi$.

If we consider a special case of $m=0$, then we can find a path satisfying \eqref{alphacond2}.
An explicit example is $p_1 = c \sin 2\theta_+$ with a positive constant $c$. %In this manner, graphene can have a new topological number for its edge state.

We conclude that the edge states of the 3D Weyl semimetals acquire a new topological charge
in the space of parameters of the boundary conditions. The 2D gapped case eliminates the
topological charge, but 2D gapless systems can have the topological charge. 
The topological charge winds the space of $\theta_+$, the only parameter dictating the
most generic boundary conditions.

%
%Suppose $m>0$, then we can see that even by tuning $p_1$, $2\theta_+$ has a limited range, if the cut-off momentum is $\Lambda$:
%\begin{align}\label{ranthe}
%	\text{arcsin}(\frac{m}{\sqrt{\Lambda^2+m^2}})<2\theta_+<2\pi-\text{arcsin}(\frac{m}{\sqrt{\Lambda^2+m^2}}),
%\end{align}
%and even with infinite cut-off, we see that, from equation \eqref{ranthe}, $2\theta_+ \neq 2\pi$. This means $2\theta_+$ cannot range from zero to $2\pi$, so topological number cannot be added up. 
%%%%%%%%%%%%%%%%%%%%%%%%%%%%
\section{Summary and discussion}

In this paper, we have studied the most general boundary conditions of the 3D Weyl semimetals
in the continuum limit around the Weyl point. The boundary conditions are shown to be dictated
by a single real parameter $\theta_+$ which takes a value on a circle in the range $0<2\theta_+\leq 2\pi$.
The edge state wave functions and their dispersion relations are obtained, and we find that
the dispersion plane as a function of the remaining conserved momenta $(p_1,p_2)$ 
terminates at the bulk dispersion cone, as a tangential plane to it. The parameter $\theta_+$
corresponds to the rotation angle of the edge dispersion plane relative to the bulk Weyl cone.

We build lattice Hamiltonian with a parameter at the boundary of the 3D square lattice, which reproduces
the $\theta_+$ dependence of the edge dispersion relations. 
Introduction of a boundary mass term at the edge of the lattice 
leads to various shape of the Fermi arcs joining Weyl nodes with opposite chiralities.
%It has been checked for various kinds of lattices and
%boundaries, such as armchair and zigzag boundaries of honeycomb lattice of graphene, and
%also on square lattices.

Through a dimensional reduction, 
the system becomes a 2D topological insulator of class A. 
The bulk-edge correspondence is found to be consistent for all values of $\theta_+$, meaning
that for any generic boundary condition the bulk-edge correspondence works.
We propose how to count the edge modes of 3D Weyl semimetals so that it becomes
consistent with the correspondence between the number of the 
edge states and the bulk topological charges.

Furthermore, we discover a new topological number for the edge states of the 3D Weyl semimetals.
The topological charge is associated with a Berry phase along the path parameterized by $\theta_+$,
the new parameter dictating the whole boundary conditions.

Various values of the  new parameter $\theta_+$ can be realized in experiments.
For example, a hydrogen termination of graphene and related materials has been studied~\cite{hattori2016edge},
and the dispersion relations obtained from the microscopic lattice Hamiltonians
exhibit a behavior which we have generically studied in this paper. To what extent our $\theta_+$ 
could be realized in experiments is one of the interesting future issues.

The meaning of the newly found topological charge needs to be verified in more details. 
The topological charge owned by the edge modes was studied in 
\cite{Hashimoto:2016dtm} for 4D topological insulators, 
inspired by a connection to superstring theory \cite{Hashimoto:2015dla}.
The edge topological charge
would indicate
existence of edge-of-edge states. For the present case, the topological charge is obtained
by the change in $\theta_+$, that is, the boundary condition itself. Therefore, to have 
such an edge-of-edge state, one needs to change $\theta_+$ as a function of the location
on the boundary. It would be quite interesting if there exists such a topologically protected
edge-of-edge state for 2D and 3D gapless systems.

In this paper our motivation came from a particle-theoretical viewpoint to explore all
possible boundary conditions in the continuum limit of the Weyl semimetals. As we summarized
above, our approach turned out to give fruitful results in condensed matter physics.
Bridges between condensed matter physics and particle physics will play crucial roles more in coming
advances in physics.

%%%%%%%%%%

\appendix

%%%%%%%%%%%

\section{Momentum-dependent generic boundary condition}\label{sec:momentum}

In this appendix, we study generic momentum-dependent boundary conditions.
We shall prove that the two properties of the edge state dispersions 
described at the end of Sec.\ref{sec:generic} 
are not modified even under a generic momentum dependence in the boundary conditions.

In the analysis so far, we have considered only a constant matrix $M$. This is because
basically we are dealing with the small momentum limit where the Weyl point is approximated
by a relativistic dispersion relation. Since the bulk Hamiltonian is linear in the momenta,
We could allow a linear dependence also for the boundary Lagrangian, which means 
$M$ linear in $p$. Since the boundary is at $x^3=0$, good quantum numbers are only
$p_1$ and $p_2$, so in $M$ let us allow a linear dependence in $p_1$ and $p_2$.
Since these are just parameters, the constraint equation for $M$ is obtained in the same 
manner, to have $A_1^2+A_2^2-B_3^2=1$. Although $A_1$, $A_2$ and $B_3$ are linear in
$p_1$ and $p_2$, this constraint equation is quadratic. This means that the linear
approximation of the boundary matrix $M$ breaks down. One needs higher order
terms in $p_1$ and $p_2$ to have a consistent boundary condition.

Let us formally continue the study of having a consistent momentum-dependent
boundary condition and allow higher order terms of the momenta in $M$.
Since we have shown above that generic boundary condition is completely dictated
by the parameter $\theta_+$, this means that the $\theta_+$ will depend on the momenta,
$\theta_+(p_1,p_2)$. Thus, we have a function-parameter family of the boundary 
condition, whose edge state has a dispersion
\begin{align}
\epsilon = - p_1 \cos (2\theta_+(p_1,p_2)) - p_2 \sin(2\theta_+(p_1,p_2)).
\label{genedge}
\end{align}

Let us show that this generalized boundary condition also shares the same properties
as that for the constant $\theta_+$. First, let us check that the edge dispersion
intersects with the bulk dispersion. The latter is a disk for a given energy $\epsilon$,
\begin{align}
\epsilon^2\geq p_1^2 + p_2^2.
\end{align}
Supposing that the boundary of the disk shares a point with the edge dispersion,
we denote it as 
\begin{align}
(p_1,p_2) = \epsilon (\cos a,\sin a). 
\label{p1p2asol}
\end{align}
The existence of $a$ means that
the edge dispersion intersects with the bulk dispersion. Substituting this expression to
the edge state dispersion
\eqref{genedge}, we find an equation for $a$,
\begin{align} 
\cos (a-2\theta_+(\epsilon \cos a, \epsilon\sin a)) = -1.
\label{consta}
\end{align}
This is equivalent to
\begin{align}
2\theta_+(\epsilon \cos a, \epsilon\sin a) = a + (2n+1)\pi, \quad n \in {\mathbb Z}.
\end{align}
Since the left hand side of this equation is a periodic function of $a$, 
we can show that this equation always have odd number of solutions
for $a$ in the period $0\leq a < 2\pi$. 
Thus the edge dispersion always intersects with the bulk dispersion.

Next, let us show that the edge dispersion is always tangential to the Weyl cone.
At the intersection point, we have a solution $a=a_0$ of the equation \eqref{consta}.
At $a=a_0$, the tangential line of the bulk dispersion at a constant 
$\epsilon$ has a slope $-\cot a_0$ in the $(p_1,p_2)$ space, 
since the Weyl cone at a slice of constant $\epsilon$ is just the disk.
Let us show that the edge dispersion \eqref{genedge} also has the same slope at
the intersection. Differentiating the dispersion relation \eqref{genedge} with respect to $p_2$,
we find
\begin{align}
\frac{dp_1}{dp_2}
=-\frac{\sin 2\theta_+}{\cos 2\theta_+}
+ 
\left( \frac{\sin 2\theta_+}{\cos 2\theta_+}p_1 -p_2\right) \frac{d(2\theta_+)}{dp_2} \, .
\end{align}
By using the intersection condition \eqref{consta} and \eqref{p1p2asol}, 
we can simplify this equation to
\begin{align}
\frac{dp_2}{dp_1} = -\cot a_0
\end{align}
which shows nothing but the slope $-\cot a_0$ in the $(p_1,p_2)$ space. 
This completes the proof that the edge dispersion is tangential to the bulk dispersion.

Finally, we show that the intersection point is in fact the point where the edge dispersion
ends. To show it, we notice that the termination condition of the edge dispersion is
$\alpha = 0$, since the existence of the edge is certified by $\alpha > 0$. Now, for 
the momentum-dependent $\theta_+$, we have the same formula
\begin{align}
\alpha = p_1 \sin \left(2 \theta_+(p_1,p_2)\right) - p_2 \cos 
\left(2\theta_+(p_1,p_2)\right). 
\end{align}
Substituting the intersection condition \eqref{p1p2asol} and \eqref{consta}, we can show 
\begin{align}
\alpha\biggm|_{a=a_0} = 0.
\end{align}
This means that the edge dispersion is actually terminated at the intersection point,
thus the edge is absorbed into the bulk.

In this manner, we can show that, even when a completely generic momentum dependence
is allowed in the boundary condition, the edge dispersion keeps the properties that it is tangential to
the bulk Weyl cone and it is terminated there.

%%%%%%%%%%%%

%%%%%%%%%%%%%%%%%%%%%%%%%%%%%%%%%%%%%

\section{Two parallel boundaries}\label{sec:2b}

In the main part of this paper, we study the general case of a single flat surface as a boundary.
For realistic materials, two parallel boundaries are typical, and in this appendix we
analyze the Weyl semimetal with two parallel boundaries, $x^3=0$ and $x^3=L$, in the continuum limit.

Each boundary can have the parameter $\theta_+$, so the number of the parameters
grows as one introduces many boundaries. In this subsection, just for simplicity, we consider the case
with identical values of $\theta_+$ for the two boundaries.

Since the bulk Hamiltonian is not altered, the energy eigen equation \eqref{HE'} does not
change, thus a generic solution is
\begin{align}
\left(
\begin{array}{c}
\xi \\ \eta 
\end{array}
\right)
= 
e^{-\alpha x^3}
\left(
\begin{array}{c}
\xi_1 \\ \eta_1 
\end{array}
\right)
+
e^{\alpha x^3} 
\left(
\begin{array}{c}
\xi_2 \\ \eta_2 
\end{array}
\right),
\end{align}
with \eqref{al}.
Note here that we need to include another mode $\exp[+\alpha x^3]$
because the allowed region of $x^3$ is a finite period $0\leq x^2 \leq L$.

Now, \eqref{HE'} leads to
\begin{align}
\left(
\begin{array}{cc}
i\alpha - \epsilon & p_1-ip_2 \\
p_1 + i p_2 & -i\alpha-\epsilon
\end{array}
\right)
\left(
\begin{array}{c}
\xi_1 \\ \eta_1 
\end{array}
\right) = 0,
\label{12-1}
\\
\left(
\begin{array}{cc}
-i\alpha - \epsilon & p_1-ip_2 \\
p_1 + i p_2 & i\alpha-\epsilon
\end{array}
\right)
\left(
\begin{array}{c}
\xi_2 \\ \eta_2 
\end{array}
\right) = 0.
\label{12-2}
\end{align}
On the other hand, the boundary condition at $x^3=0$ and $x^3=L$ leads to
\begin{align}
(1 \;\; e^{-2i\theta_+})
\left(
\begin{array}{c}
\xi_1+\xi_2 \\ \eta_1+\eta_2 
\end{array}
\right) = 0, \\
(1 \; \; e^{-2i\theta_+})
\left(
\begin{array}{c}
e^{-\alpha L}\xi_1+e^{\alpha L}\xi_2 \\ 
e^{-\alpha L}\eta_1+e^{\alpha L}\eta_2 
\end{array}
\right) = 0
\end{align}
which are equivalent to
\begin{align}
(1 \;\; e^{-2i\theta_+})
\left(
\begin{array}{c}
\xi_1 \\ \eta_1 
\end{array}
\right) = 
(1 \; \; e^{-2i\theta_+})
\left(
\begin{array}{c}
\xi_2 \\ 
\eta_2 
\end{array}
\right) = 0 \, .
\end{align}
So, we can solve the wave function as if two boundaries are just a set of copies of
a single boundary, thanks to our assumption that the same boundary conditions are shared
by the two.
Together with \eqref{12-1} and \eqref{12-2}, we have zero-mode equations
\begin{align}
\left(
\begin{array}{cc}
i\alpha - \epsilon & p_1-ip_2 \\
1 & e^{-2i\theta_+}
\end{array}
\right)
\left(
\begin{array}{c}
\xi_1 \\ \eta_1 
\end{array}
\right) = 0,
\label{12-3}
\\
\left(
\begin{array}{cc}
-i\alpha - \epsilon & p_1-ip_2 \\
1 & e^{-2i\theta_+}
\end{array}
\right)
\left(
\begin{array}{c}
\xi_2 \\ \eta_2 
\end{array}
\right) = 0.
\label{12-4}
\end{align}
If both of these have nontrivial eigenvector solutions, it leads to
\begin{align}
(i\alpha -\epsilon)e^{-2i\theta_+}  
= 
(-i\alpha -\epsilon)e^{-2i\theta_+} = p_1-ip_2.
\end{align}
However this equation has no solution for generic $p_1$ and $p_2$. Therefore,
either \eqref{12-3} or \eqref{12-4} is solved by a trivial vanishing solution.

When \eqref{12-3} has a nontrivial solution, $\xi_2=\eta_2=0$, then the wave function
and the energy eigenvalue is completely identical to the previous case of a single
boundary:
\begin{align}
\epsilon = -p_1 \cos 2\theta_+ - p_2 \sin 2\theta_+ ,
\label{ep2} \\
\alpha = p_1 \sin 2\theta_+ - p_2 \cos 2\theta_+ .
\label{al2}
\end{align}
This is the edge state dispersion, which turn out to be identical to 
the single boundary case, \eqref{rotea}.

On the other hand, when \eqref{12-4} has a nontrivial solution, then
$\xi_1=\eta_1=0$ and we obtain another edge state
\begin{align}
\epsilon = -p_1 \cos 2\theta_+ - p_2 \sin 2\theta_+ , \\
\alpha = -p_1 \sin 2\theta_+ + p_2 \cos 2\theta_+ .
\end{align}
This expression differs by just a sign of $\alpha$, compared to the previous one.
Furthermore, since the front factor is $\exp[\alpha x^3]$ which differs
only by the sign of $\alpha$ compared to the previous one, this solution
turns out to be identical to the previous one. So, we conclude that we have a single edge state
given by \eqref{ep2} and \eqref{al2}.

The only difference compared to the single boundary case is that now there is no
restriction $\alpha>0$. In fact, depending on the momenta $(p_1,p_2)$, 
$\alpha$ can be positive or negative, and depending on it, whether the wave function
localizes at $x^3=0$ or $x^3=L$ will be determined. Thus the obtained edge state,
although written in a unified form, represents both the modes, one localized at $x^3=0$
and the other at $x^3=L$.

\vspace{10mm}

%%%%%%%%%%%%%%%%%%%%%%%%%%%%%%%%%%%%%
\begin{acknowledgments}
We would like to thank valuable discussions with M.~Furuta, Y.~Hatsugai and T.~Fukui.
The work of K.~H.~was supported in part by JSPS KAKENHI
Grant Number JP15H03658 and JP15K13483.
The work of TK was supported in part by Keio Gijuku Academic Development Funds, the MEXT-Supported Program for the Strategic Research Foundation at Private Universities ``Topological Science'' (No. S1511006), and JSPS Grant-in-Aid for Scientific Research on Innovative Areas ``Topological Materials Science'' (No. JP15H05855).
\end{acknowledgments}

%%%%%%%%%%% References %%%%%%%%%%%%%%%%%%%%%%%%%
%\newcommand{\J}[4]{#1 {\bf #2} (#3) #4}
%\newcommand{\andJ}[3]{{\bf #1} (#2) #3}
%\newcommand{\AP}{Ann.\ Phys.\ (N.Y.)}
%\newcommand{\MPL}{Mod.\ Phys.\ Lett.}
%\newcommand{\NP}{Nucl.\ Phys.}
%\newcommand{\PL}{Phys.\ Lett.}
%\newcommand{\PR}{ Phys.\ Rev.}
%\newcommand{\PRL}{Phys.\ Rev.\ Lett.}
%\newcommand{\PTP}{Prog.\ Theor.\ Phys.}
%\newcommand{\hep}[1]{{ hep-th/{#1}}}
%%%%%%%%%%%%%%

%\bibliographystyle{ytphys}
\bibliography{3dTI-revtex}

\end{document}